\documentclass[10pt,letterpaper]{IEEEtran}
\usepackage{amsmath,amssymb,amsfonts,amsthm}
\usepackage{bbm}

\usepackage{gensymb}%degree symbol
\usepackage{graphicx}
\usepackage{subfigure}
\usepackage[dvipsnames,usenames]{color}
\usepackage{hyperref}
\hypersetup{
	colorlinks=true,
	linkcolor=blue,
	filecolor=blue,
	citecolor = red,      
	urlcolor=cyan,
}

\usepackage[lmargin=0.755in,rmargin=0.755in,tmargin=0.755in,bmargin=0.755in]{geometry}

\usepackage{nomencl}
\makenomenclature

\usepackage[numbers,sort&compress]{natbib}

\newtheorem{remark}{Remark}

\newtheorem{prop}{Proposition}

\usepackage{adjustbox}
\newcommand{\markedManu}{MARKED} % MARKED, unMarked
\ifthenelse{\equal{\markedManu}{MARKED}}
{

}

\ifthenelse{\equal{\markedManu}{nonMARKED}}
{

}

\def\matlab{Matlab}

\def\rd#1{{\color{red}{#1}}}

\def\R{\mathbb{R}}

\def\eqdef{\ensuremath{:=}}
\def\UB{\text{UB}}
\def\LB{\text{LB}}

\def\plantFullName{Central Chilled Water Plant}
\def\plant{CCWP}
\def\plants{CCWPs}
\def\tot{\text{tot}} % required
\def\req{\text{req}} % required
\def\set{\text{set}} % required

\def\pw{{\text{pw}}} % for specific heat of water , C_\pw

\def\noml{\text{nom}}

\def\lw{\text{lw}}
\def\lws{\text{lws}}
\def\lwr{\text{lwr}}
\def\cc{\text{cc}} % cooling coil
\def\sw{\text{sw}} %supply water: after the bypass volve but before...
\def\rw{\text{rw}}
\def\twc{\text{twc}}
\def\tww{\text{tww}}
\def\tw{\text{tw}}
\def\tes{\text{tes}}
\def\evap{\text{evap}}
\def\ch{\text{ch}}
\def\chw{\text{chw}}
\def\chws{\text{chws}}
\def\chwr{\text{chwr}}
\def\cond{\text{cd}}
\def\condwr{\text{cdwr}}
\def\bp{\text{bp}} % bypass
\def\chwp{\text{chwp}}

\def\ct{\text{ct}}
\def\cw{\text{cw}} 
\def\cwr{\text{cwr}}
\def\cws{\text{cws}}
\def\wat{\text{wat}}
\def\air{\text{air}}
\def\ran{\text{ran}}
\def\app{\text{app}}
\def\sco{\text{sco}}
\def\oawb{\text{oawb}}
\def\oa{\text{oa}}
\def\cwp{\text{cwp}}

\def\qDot{\ensuremath{\dot{q}}}
\def\qDotL{\ensuremath{\dot{q}^{\text{L}}}}
\def\qDotCC{\ensuremath{\dot{q}^{\cc}}}
\def\qDotEvap{\ensuremath{\dot{q}^{\evap}}}
\def\qDotEvapNoml{\ensuremath{\dot{q}^{\evap}_{\noml}}}

\def\qDotEvapi{\ensuremath{\dot{q}^{\evap,i}}}

\def\qDotCond{\ensuremath{\dot{q}^{\cond}}}

\def\qDotCondi{\ensuremath{\dot{q}^{\cond,i}}}
\def\qDotCT{\ensuremath{\dot{q}^\mathrm{ct}}}
\def\qDotCTi{\ensuremath{\dot{q}^\mathrm{ct,i}}}

\def\PCh{\ensuremath{P^{\ch}}}

\def\PChNoml{\ensuremath{P^{\ch}_{\noml}}}

\def\PChi{\ensuremath{P^{\ch,i}}}

\def\PCT{\ensuremath{P^{\ct}}}
\def\PCTNoml{\ensuremath{P^{\ct}_{\noml}}}

\def\PCTi{\ensuremath{P^{\ct,i}}}

\def\PChWP{\ensuremath{P^{\chwp}}}
\def\PChWPi{\ensuremath{P^{\chwp,i}}}
\def\PCWP{\ensuremath{P^{\cwp}}}
\def\PCWPi{\ensuremath{P^{\cwp,i}}}
\def\Ptot{\ensuremath{P^{\tot}}}

\def\CapFunT{\ensuremath{f^{\text{CapFunT}}}}
\def\EIRFunT{\ensuremath{f^{\text{EIRFunT}}}}
\def\EIRFunPLR{\ensuremath{f^{\text{EIRFunPLR}}}}
\def\CR{\text{CR}}

\def\TLWS{\ensuremath{T^{\lws}}}
\def\TLWR{\ensuremath{T^{\lwr}}}
\def\TTWC{\ensuremath{T^{\twc}}}
\def\TTWW{\ensuremath{T^{\tww}}}
\def\TChWS{\ensuremath{T^{\chws}}}
\def\TChWSSet{\ensuremath{T^{\chws}_{\set}}}

\def\TChWSi{\ensuremath{T^{\chws,i}}}

\def\TChWR{\ensuremath{T^{\chwr}}}
\def\TChWRi{\ensuremath{T^{\chwr,i}}}
\def\TCWS{\ensuremath{T^{\cws}}}
\def\TCWSi{\ensuremath{T^{\cws,i}}}

\def\TCWR{\ensuremath{T^{\cwr}}}
\def\TCWRi{\ensuremath{T^{\cwr,i}}}
\def\TCWRUB{\ensuremath{T^{\cwr}_{\UB}}}
\def\TCONDWRUB{\ensuremath{T^{\condwr}_{\UB}}}

\def\TOAWB{\ensuremath{T^{\text{oawb}}}}
\def\TTWC{\ensuremath{T^{\twc}}}
\def\TSW{\ensuremath{T^{\sw}}}
\def\TRW{\ensuremath{T^{\rw}}}
\def\TRan{\ensuremath{T^{\ran}}}
\def\TApp{\ensuremath{T^{\app}}}

\def\mDot{\ensuremath{\dot{m}}}
\def\mDotLW{\ensuremath{\dot{m}^{\lw}}}
\def\mDotTW{\ensuremath{\dot{m}^{\tw}}}
\def\mDotSW{\ensuremath{\dot{m}^{\sw}}}
\def\mDotBP{\ensuremath{\dot{m}^{\bp}}}
\def\mDotCHW{\ensuremath{\dot{m}^{\chw}}}
\def\mDotCHWOne{\ensuremath{\dot{m}^{\chw,1}}}

\def\mDotCHWi{\ensuremath{\dot{m}^{\chw,i}}}
\def\mDotCOND{\ensuremath{\dot{m}^{\cond}}}

\def\mDotCONDOne{\ensuremath{\dot{m}^{\cond,1}}}

\def\mDotCONDi{\ensuremath{\dot{m}^{\cond,i}}}

\def\TCONDWS{\ensuremath{T^{\mathrm{cdws}}}}
\def\TCONDWR{\ensuremath{T^{\mathrm{cdwr}}}}
\def\TCONDWSi{\ensuremath{T^{\mathrm{cdws},i}}}
\def\TCONDWRi{\ensuremath{T^{\mathrm{cdwr},i}}}

\def\mDotCW{\ensuremath{\dot{m}^{\cw}}}
\def\mDotCWNoml{\ensuremath{\dot{m}^{\cw}_{\noml}}}

\def\mDotCWi{\ensuremath{\dot{m}^{\cw,i}}}

\def\mDotOA{\ensuremath{\dot{m}^{\oa}}}
\def\mDotOANoml{\ensuremath{\dot{m}^{\oa}_{\noml}}}

\def\STWC{\ensuremath{S^{\twc}}}
\def\STWW{\ensuremath{S^{\tww}}}
\def\STW{\ensuremath{S^{\tw}}}
\def\MTES{\ensuremath{M^{\tes}}}

\def\YorkCalc{\texttt{YorkCalc}}
\def \YorkCalcOurs{\texttt{YorkCalc:wSat}}
\def \TRanState{\ensuremath{T^{\mathrm{ran}}}}
\def \TAppState{\ensuremath{T^{\mathrm{app}}}}

\def\nCh{\ensuremath{n^{\ch}}}
\def\nCT{\ensuremath{n^{\ct}}}

\def\PLR{\ensuremath{\mathrm{PLR}}}

        \def\TCWRav{\ensuremath{\bar{T}^{\mathrm{cwr}}}}
        \def\TCWSav{\ensuremath{\bar{T}^{\cws}}}
        \def\TChWSav{\ensuremath{\bar{T}^{\chws}}}
\def\TChWRav{\ensuremath{\bar{T}^{\chwr}}}

\def\ElectricEIR{\texttt{Electric:EIR}}
\def\OurElectricEIR{\texttt{Electric:EIR:wSat}}

\def\rd#1{{\color{red}{#1}}}

\def\pb#1{\footnote{pb: \rd{#1}      }}

\newlength{\noteWidth}
\long\def\notes#1{\ifinner
	{\footnotesize #1}
	\else
	\marginpar{\parbox[t]{\noteWidth}{\raggedright\footnotesize #1}}
	\fi\typeout{#1}}

\usepackage{etoolbox}
\renewcommand\nomgroup[1]{%
	\item[\bfseries
	\ifstrequal{#1}{A}{Variables}{%
		\ifstrequal{#1}{B}{Superscripts}{%
			\ifstrequal{#1}{C}{Subscripts}{%
				\ifstrequal{#1}{D}{Others}{}}}}%
	]}

\usepackage{xparse}
\ExplSyntaxOn%% allow _ : as a letter

\ExplSyntaxOff
\ExplSyntaxOn
\cs_new:Nn \pb_prop_gset_bykeys:Nn
{
	\prop_clear:N \l__pb_temp_prop
	\keys_set:nn { pb/propbykey } { #2 }
	\prop_gset_eq:NN #1 \l__pb_temp_prop
}
\cs_generate_variant:Nn \pb_prop_gset_bykeys:Nn { c }

\keys_define:nn { pb/propbykey }
{
	unknown .code:n = \prop_put:Nxn \l__pb_temp_prop { \l_keys_key_tl } { #1 }
}

\cs_generate_variant:Nn \prop_put:Nnn { Nx }

\NewDocumentCommand{\setupcollaborator}{mm}
{% #1 = identifier string, #2 = set of key-value pairs
	\prop_new:c { g_collaborator_#1_prop }
	\pb_prop_gset_bykeys:cn { g_collaborator_#1_prop } { #2 }
}

\NewDocumentCommand{\selectcollaborator}{m}
{
	\prop_map_inline:cn { g_collaborator_#1_prop }
	{
		\tl_set:cn { ##1 } { ##2 }
	}
}
\ExplSyntaxOff

\setupcollaborator{PBmacbookairTIDF}
{% pb, macbook air or mac mini at work, 
	NAME=Prabir Mac , laptop or mac mini,
	DiCEbibPATH=/Users/pbarooah/Library/CloudStorage/Dropbox/DiCElab_bib,
}
\setupcollaborator{PBmacbookairPersonal}
{% pb, macbook air or mac mini at work, 
	NAME=Prabir Macbook air, personal, 2022,
	DiCEbibPATH=/Users/prabirbarooah/bibfilesFromDropbox/DiCElab_bib,
	JBbibPATH=/Users/prabirbarooah/bibfilesFromDropbox/bibs,
	ACbibPATH=/Users/prabirbarooah/bibfilesFromDropbox/austin_bib, 
}
\setupcollaborator{GZlinux}
{% GZlinux is using his lab computer
	NAME=GZ on his lab computer,
	DiCEbibPATH=../../DiCElab_bib,
}

\selectcollaborator{PBmacbookairTIDF}

\begin{document}
	\title{\vspace{0.25in} \centering A
          Central Chilled Water Plant Model for Designing Learning-Based
          Controllers}
	\author{
		\IEEEauthorblockN{Zhong	Guo\IEEEauthorrefmark{1} and Prabir Barooah \IEEEauthorrefmark{2} \\\IEEEauthorrefmark{1}University of Florida \\\IEEEauthorrefmark{2}Indian Institute of Technology, Guwahati} 
		\thanks{\IEEEauthorrefmark{2} Corresponding author, email: pbarooah@iitg.ac.in.}
		\thanks{Both authors were with the Dept. of Mechanical and
                  Aerospace Engineering, University of Florida (UF),
                  Gainesville, FL 32601, USA. PB is currently with the
                  Dept. of Electronics and Electrical Engineering,
                  Indian Institute of Technology, Guwahati, Assam,
                  781039, India. This work was supported by the
                  National Science Foundation, USA under grants
                  1934322 (CMMI) and 2122313 (ECCS) and  the Science and Engineering Research Board of India,
                  under grant CRG/2023/008102. }
		\vspace{-0.25cm}
	}
	\maketitle
	\begin{abstract}
		We describe a framework of modeling a central chilled
                water plant (\plant) that consists of an aggregate
                cooling coil, a number of heterogeneous chillers and
                cooling towers, and a chilled water-based thermal
                energy storage system. We improve upon existing component
                models from the open literature using a constrained
                optimization-based framework to ensure that the models
                respect capacities of all the heat exchangers (cooling
                coils, chillers, and cooling towers) irrespective of
                the inputs provided. As a result, the proposed model has a wider
                range of validity compared to existing models; the
                latter can                produce highly erroneous outputs when inputs are not
                within normal operating range. This
                feature is essential for training learning-based 
                controllers that can choose inputs beyond normal operating conditions and is lacking in currently available
                models. The overall plant model is
                implemented in Matlab and is made publicly
                available. Simulation of a \plant\ with closed loop
                control is provided as an illustration.
	\end{abstract}
	
	\nomenclature[A]{$T$}{Temperature (Celsius)}
	\nomenclature[A]{\mDot}{Flowrate (kg/sec)}
	\nomenclature[A]{\qDot}{Heat Exchange Rate (kW)}
	\nomenclature[A]{$P$}{Electric Power Consumption (kW)}
	\nomenclature[A]{$S$}{Fraction of Water in TES}
	
	\nomenclature[B]{L}{Load}
	\nomenclature[B]{\cc}{Cooling Coil}
	\nomenclature[B]{\tes}{Thermal Energy Storage}
	\nomenclature[B]{\tww}{Warm TES sub-tank}
	\nomenclature[B]{\twc}{Cold TES sub-tank}
	\nomenclature[B]{\ch}{Chiller}   
	\nomenclature[B]{\chw}{Chilled Water}
	\nomenclature[B]{\chwr}{Chilled Water Return}
	\nomenclature[B]{\chws}{Chilled Water Supply}
	\nomenclature[B]{\evap}{Chiller Evaporator}
	\nomenclature[B]{\cond}{Chiller Condenser}
	\nomenclature[B]{\sw}{Chilled Water Loop Supply Water}
	\nomenclature[B]{\rw}{Chilled Water Loop Return Water}
	\nomenclature[B]{\ct}{Cooling Tower}
	\nomenclature[B]{\cw}{Cooling Water}
	\nomenclature[B]{\cwr}{Cooling Water Return}
	\nomenclature[B]{\cws}{Cooling Water Supply}
	\nomenclature[B]{\oa}{Outside Air}
	\nomenclature[B]{\chwp}{Chilled Water Pump}
	\nomenclature[B]{\cwp}{Cooling Water Pump}
	
	\nomenclature[C]{$k$}{Timestep Index}
	\nomenclature[C]{\LB/\UB}{Lower/Upper Bound}
	\nomenclature[C]{\req}{Required}
	\nomenclature[C]{\set}{Setpoint}
	\nomenclature[C]{\noml}{Nominal} 
	
	\nomenclature[D]{\MTES}{Total Water Mass in TES (kg)}
	\nomenclature[D]{MBL}{Modelica Buildings Library}
	
	\printnomenclature
	\section{Introduction}
Central chilled water plants (\plants) consist of chillers, cooling
towers, and increasingly thermal energy storage  (TES) systems.  They
are variously called ``chiller plants'', ``central plants'',
``district cooling energy plants'', etc. \plants\ are widely used
in campuses to provide cooling to a cluster of
buildings. Optimal
control of \plants\ have attracted quite a bit of research. Apart from
model-based optimal control such as MPC~\cite{Huang_BldgAndEnv2017,DengMPCASE:2014,SchweigerDistrictEnergy:2017,Patel_IntlHPB2018_HVACMPC,ZabalaVirtualRSER:2020},
there has recently been considerable interest in learning-based control
methods, especially reinforcement
learning~\cite{ParkMachineSTBE:2019,QiuModelfreeSTBE:2020,GuoReinforcementACC:2022,WongOptimizingArXiV:2022,RosdahlModel-freeSTBE:2023}
and extremum-seeking~\cite{MuRealtimeAE:2017,VuEnergy:HPB:2018}.

Learning-based model-free controllers, such as reinforcement learning
(RL) and extremum seeking controllers, need to explore the action
space to learn the optimal control policy. Such exploration can drive
the system outside its normal and safe operating
conditions. Additionally, training such a controller for a complex system like a \plant\ typically requires a long and costly training phase.
Due to the risk and cost of learning directly from experiments
conducted on a physical \plant, these ``model-free''
controllers are typically designed by conducting numerical experiments on a realistic simulator. There
has been substantial work on modeling \plants\ and their
components - chillers, cooling towers and TES systems - spanning decades;
see~\cite{ASHRAE_handbook_equipment:20} and references therein. Some of the well-established
component models are now incorporated into EnergyPlus~\cite{EnergyPlus:00} and the Modelica Buildings Library (MBL)~\cite{ModelicaBldgLibrary:2014}. There are works specifically on modeling \plants\ by combining existing equipment models, and controller development is an usual motivation behind these works~\cite{BaillieDevelopmentATE:2017,JaramilloPHDthesis:2018, FanOpensourceAE:2021}, as it is in our case.

However, there is currently a lack of easy-to-use open-source models
of~\plants\ that can be used for training learning-based controllers. One of the
main reasons is the \emph{limited region of validity} of these
models. They often produce poor predictions when inputs are chosen
outside normal operation range. For studies with classical control
algorithms, this feature is not a serious hurdle since control
commands can be chosen to be within normal operating range. However, the
situation is quite different in case of an RL controller. An RL controller
implicitly learns system constraints by getting poor rewards when its
actions lead to constraint violations, such as a lack of adequate
cooling to buildings. Therefore the simulation model must allow operation outside normal
operating conditions and predict poor performance of the plant therein. For example, if a small condenser water flow
rate is used under a high cooling load condition, the model must
predict that the plant is unable to reject the heat from the buildings
and the cooling load is not serviced. This is also true for
extremum seeking, though to a lesser degree. However, existing models of
chillers and cooling towers, and thus of \plants\ composed of them,
can predict  utterly unreasonable outputs 
when the inputs are outside the designed operating conditions. We show an example of
such behavior - limited region of validity - of a commonly used chiller model in Sec.~\ref{sec:OriginalChillerFail}.

\begin{figure*}[h]
		\centering
		\includegraphics[width=2\columnwidth,clip=true,
                trim=0.5in 3.4in 0.5in 2.2in]{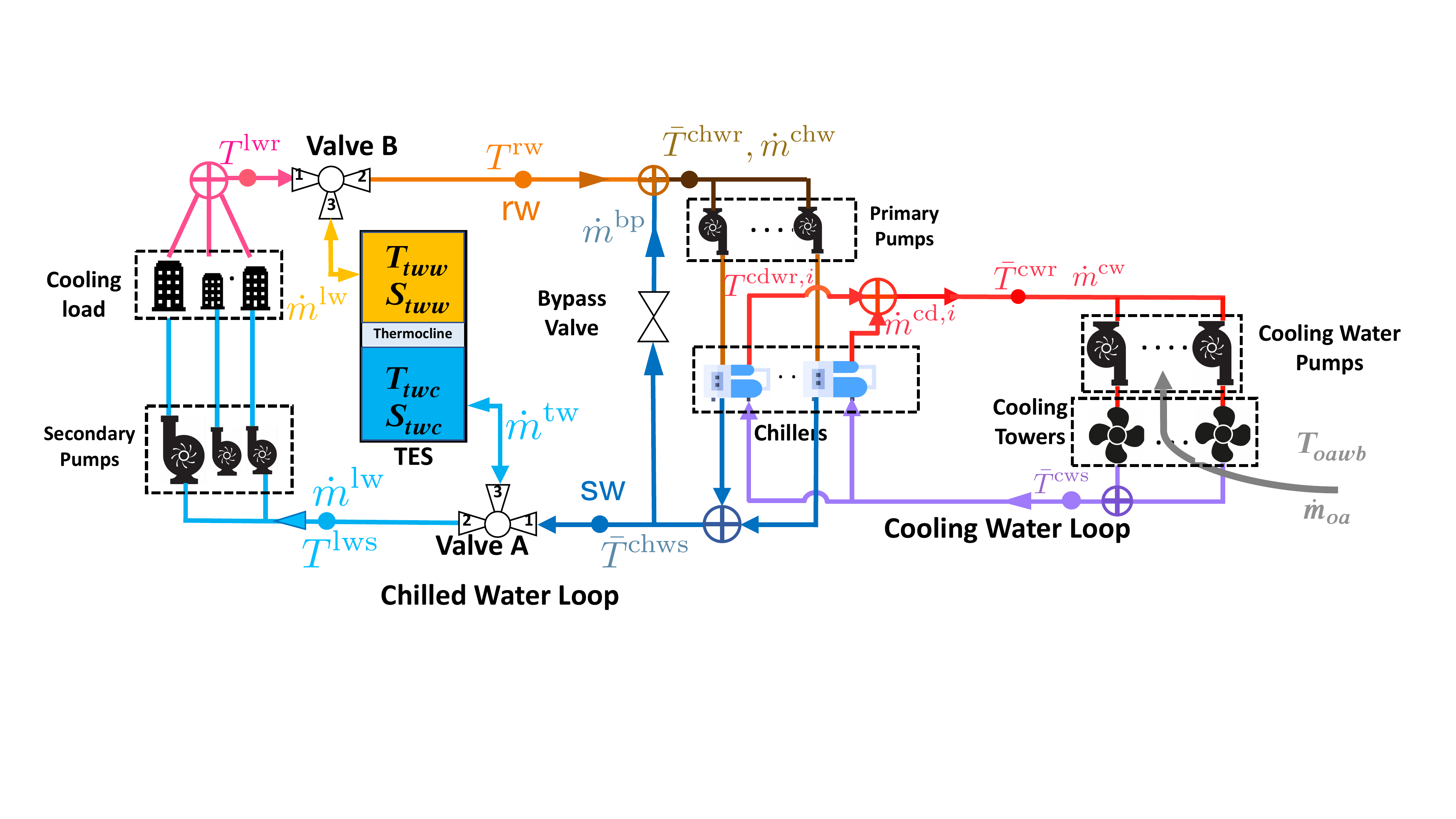}
		\caption{Configuration of a \plantFullName\ considered
                  in this paper.}
		\label{fig:DetailCCWP}
	\end{figure*}

This paper presents an open-source model for the simulation of
\plants\ that fills the gap mentioned above. The schematic of the
\plant\ for which the proposed model is developed is shown in
Fig.~\ref{fig:DetailCCWP}. The components of the central plant (cooling coil, TES, chiller, and cooling tower) are modeled by
modifying existing models from the open literature, some of which are already
part of EnergyPlus and MBL. Our key innovation is to formulate some of the
equipment models as solutions to constrained
optimization problems so that the capacities of all the heat exchangers
(the cooling coil, chillers, and cooling towers) are respected. As a
result, the proposed \plant\ model is able to generate reasonable
outputs when inputs are chosen outside the designed operating
conditions of the \plant. Matlab implementation is available in \url{https://gitlab.com/pbarooah/ccwp_model}.

The rest of the paper is organized as
follows. Section~\ref{sec:OriginalChillerFail} provides evidence of
the limited range of validity of an existing chiller model, which is the motivation behind
the paper. Section~\ref{sec:contribution} summarizes contributions
over prior art. The model of each equipment in a \plant\ is described
in Section~\ref{sec:equip}. The interconnection of these component
models to construct the complete model of a \plant\ is described in
Section~\ref{sec:FullModel}. Simulations of an illustrative \plant\
model are reported in Section~\ref{sec:sim}. The paper concludes with some comments on future work in Section~\ref{sec:conclusion}.

        \subsection{Evidence of limited region of validity}\label{sec:OriginalChillerFail}
	The \ElectricEIR\ model of a vapor compression
        centrifugal chiller has been widely used in many
        studies~\cite{OpenBuildingControl,IBPSA_Project1}. It is
        currently available in
        EnergyPlus~\cite{DOEEnergyPlusEngRef:2022} and the
        MBL~\cite{ModelicaBldgLibrary:2014}. In this section we
        present a scenario in which the existing \ElectricEIR\
        model predictions are dramatically incorrect when subjected to
        inputs that are outside normal operating conditions. This
        failure is the motivation for the new approach to modeling
        proposed in this paper. 
	
\begin{figure}[t]
		\centering		
		\subfigure[junk][Testing environment for an \ElectricEIR\ chiller in Dymola.
		\label{fig:ChillerDymola}]
		{
			\includegraphics[width=0.8\columnwidth]{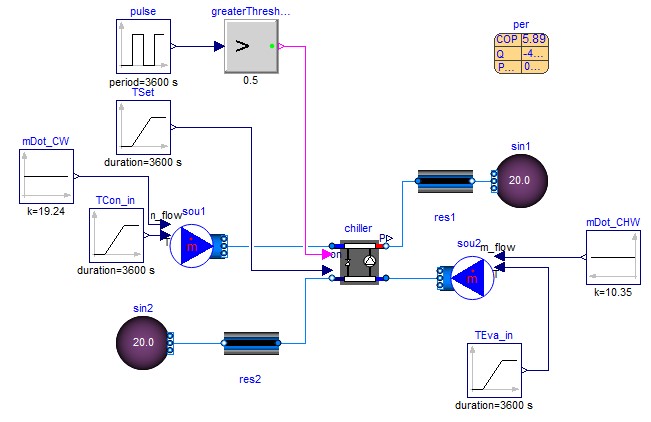}
		}
		\subfigure[junk][Prediction of cooling water return
                temperatures for various condenser water flow
                rates.
		\label{fig:CoolingWaterBlowup}]
		{
			\includegraphics[width=0.8\columnwidth]{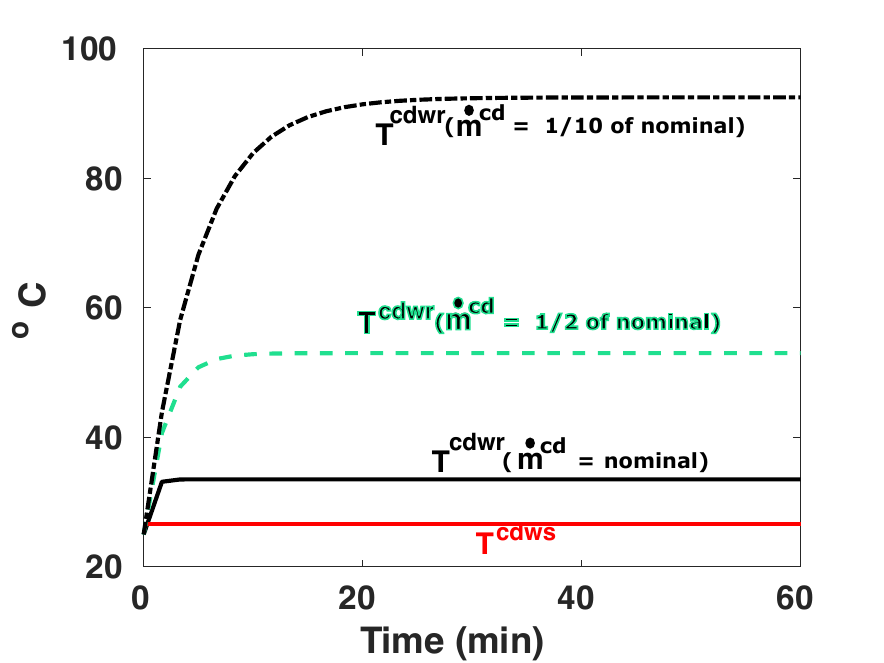}
		} 
		\caption{Evidence of limited region of validity of the
                  \ElectricEIR\ chiller model: extremely low condenser
                  water supply flow rate incorrectly predicts an extremely high cooling water return
                  temperature.} 
		\label{fig:CoolingWaterBlowup-combined}
	\end{figure}

	An \ElectricEIR\ chiller from the MBL is put into a
        simulation test environment where all inputs to a chiller can
        be varied; see Figure~\ref{fig:ChillerDymola}. The chiller has
        a rated capacity of 471.2 kW, and is of type
        \texttt{McQuay\_WSE\_471kW\_5\_89COP\_Vanes}. When the inputs
        are at their nominal values, the output -  condenser
        water return temperature, \TCONDWR\ - is at its nominal value, $35\degree$C; see
        Figure~\ref{fig:CoolingWaterBlowup}.  As the condenser water
        flow rate $\mDotCOND$ is decreased, the predicted cooling water return temperature
        increases. When the cooling water flow rate is reduced to one
        tenth of its nominal value, the model predicts that  $\TCONDWR$ is $90\degree$C in steady state;
        see Figure~\ref{fig:CoolingWaterBlowup}. This is
        clearly incorrect; in reality no chiller would exhibit a
        cooling water return temperature close to the boiling
        point. In fact, the condenser return water temperature should be below $48.9\degree$C to prevent damaging water pipes and cooling towers~\cite{Hensley_CoolingTower1985}.

	The culprit is the lack of a hard upper bound on the condenser
        capacity in the \ElectricEIR\ chiller model. The model assumes that the
        cooling load is transferred to the condenser water by the refrigerant no matter how
        hot the water stream becomes. In reality,
        physics imposes \emph{saturation}: if the heat exchange
        capacity of the equipment is reached, the outlet temperature
        will not increase further no matter the
        input is. The existing \ElectricEIR\ model fails to capture this behavior. The underlying chiller model in the MBL is the same as that in EnergyPlus. 

	\subsection{Contribution Over Prior Art}\label{sec:contribution}
Our first contribution is a novel method to ensure that the
state-dependent capacities of all the heat exchangers - cooling coils
in air handling units, evaporator and condenser in chillers, and cooling towers - are all respected at each simulation timestep. While there are research papers that have addressed
this issue for some equipment, such as~\cite{Liu_EnerAndBldg2017},
those ideas have not been translated into these open-source simulation
tools. The approach we use in this paper to enforce hard limits on heat exchanger capacities is to formulate the state propagation problem - to compute next state from current states and inputs -  as a constrained optimization problem: the actual cooling \emph{delivered} is one that is as close as possible to the cooling \emph{demand} while staying within the heat exchanger capacities.  The cooling demand is specified as problem data. Due to the availability of modern optimization modeling tools, specifying and solving the optimization problem is straightforward. This approach is inspired partially by the
iterative computations used in EnergyPlus and MBL models as well as in
the method proposed in~\cite{Liu_EnerAndBldg2017}. The underlying NLP solver uses an iterative scheme to find an optimal solution satisfying the constraints when it exists, but using modern tools for optimization makes it far easier to formulate and solve the constrained optimization problem than implementing an iterative scheme manually.

Our second contribution is that a \matlab-based software
implementation of the proposed model is made publicly available
(\url{https://gitlab.com/pbarooah/ccwp_model}). This software can be used by researchers to test and compare multiple control
methods.

Our proposed model structure is divided into three
levels: equipment, water loop, and plant. Connecting equipment models
belonging to the same water loop results in the corresponding water
loop model and combining multiple water loops leads to the plant
model. This modular framework greatly reduces the effort for building
a customized \plant\ model. For example, swapping a chiller only
changes the equipment level program; adding chillers or implementing
new pump strategies only changes the water loop level program.

A preliminary version of the model developed here was described in~\cite{GuoReinforcementACC:2022,GuoOptimalJESBC:2024}. The model in \cite{GuoReinforcementACC:2022,GuoOptimalJESBC:2024} was limited to homogeneous chillers and cooling towers, while in this paper each cooling tower and each chiller has a distinct sub-model so that heterogeneous equipment can be modeled in the proposed framework. Another major difference is that in~\cite{GuoReinforcementACC:2022,GuoOptimalJESBC:2024}, the plant model was not modular; in fact both supervisory controller computation and the plant's state propagation was cast as one single constrained optimization problem. In contrast, each equipment sub-model's computations in this work is separate, which makes the overall plant model to be modular and also provides a clear separation between the model's state propagation and supervisory controller's computations.

	\section{Equipment Models} \label{sec:equip}       
We start with a few preliminaries and then describe the model of each
equipment in a \plant. There are five major equipments in a \plant:
the cooling coil, the thermal energy storage (TES), the chiller(s),
the cooling tower(s), and various water pumps: chilled water pumps at
the chillers and the cooling water pumps at the cooling towers. For
each equipment model, we list the control input 
$u$, disturbance input $w$, and state $x$. All equipment models are discrete time dynamical systems of
the form: $x_{k+1} = f(x_k, u_k, w_k)$ and $y_k = g(x_k,
u_k)$, with $k$ being the discrete time index. The sampling period is denoted by $t_s$.

	\subsection{Preliminaries} \label{sec:prelim}
	\subsubsection{Variable naming convention}
	Variables are named with the following convention: \emph{the
          supply temperature is (nominally) lower than the return
          temperature}. Since the water returning from one equipment
        is often supplied to another equipment, such as cooling coil
        to chiller, the words ``supply'' and ``return'' can be
        confusing without providing sufficient context to indicate
        whether it is  ``supplied by'' or ``supplied to''. Providing
        such meticulous context becomes cumbersome, and requires too
        many super/sub-scripts. The convention we follow here is
        consistent and straightforward. Note that heating equipment such as boilers are not
        considered here, which may require a revision of this convention.
	
	\subsection{Aggregate Cooling Coil Model}
\begin{figure}
		\centering
		\includegraphics[width=0.9\columnwidth]{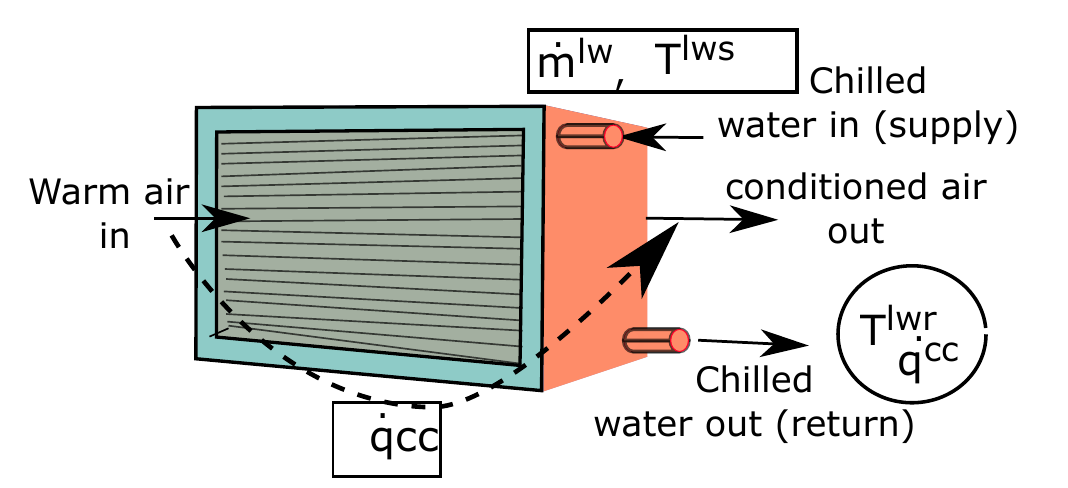}
		\caption{Aggregate load/cooling coil. Model inputs (including disturbance) in rectangles and outputs in circle.}
		\label{fig:CoolingCoil}
	\end{figure}

	In the hydronic cooling system considered in
        Figure~\ref{fig:DetailCCWP}, each building has one or more air
        handing units (AHUs). In each AHU, there is a cooling coil in
        which the chilled water supplied from chiller(s) is used to
        cool and dehumidify air before the air is resupplied to
        buildings' interior; see Fig.~\ref{fig:CoolingCoil}. Since air-side modeling is not
        a focus of this paper, we propose an \emph{aggregate} cooling
        coil model that predicts the aggregate behavior of all the
        cooling coils combined. The superscript ``lw'' - for
        ``load water'' - to denote water supplied to the
        cooling load (the aggregate of all AHUs) by the aggregate coil. The control input $u^{\cc}$, 
        disturbance $w^{\cc}$, and state $x^{\cc}$ of the model are:
	\begin{align}
		u^{\cc} &= [\TLWS, \mDotLW]^T, \label{eq:uCC}\\
	x^{\cc} &= \TLWR,  \label{eq:xCC} \\
		w^{\cc} &=  \qDotL,\label{eq:wCC} 
	\end{align}
	where $\TLWS$ and $\TLWR$ are temperatures of the chilled
        water at inlet (supply) and outlet (return) of the cooling coil; $\mDotLW$ is
        the flowrate of chilled water through the cooling coil;
        $\qDotL$ is the cooling required by the aggregate load, and $\qDotCC$ is the actual cooling provided
        by the cooling coil, i.e., heat absorbed from the air stream
        by the chilled water stream; Heat balance across the water stream, with a low-pass
        filter to capture the time constant of the heat exchange,
        gives us
	\begin{align}
		\TLWR_{k+1} &= a^{\cc}\TLWR_{k} + (1-a^{cc})\frac{1}{C_{\pw}\mDotLW_k}(\qDotCC_k + C_{\pw}\mDotLW_k \TLWS_k),\label{eq:cc-TLWR}
	\end{align}
	where $a^{\cc}= e^{-Ts/\tau}$ is a low pass filter coefficient
        with $\tau$ being the time constant of the cooling coil, $C_{\pw}$ is the specific heat of water, and
	\begin{align}\label{eq:cc-qdotcc}
		\qDotCC_{k} & =                          \min(\qDotL_k,\ \qDotCC_{\UB,k}),
	\end{align}
in which $\qDotCC_{\UB,k}$ is the maximum possible heat exchange rate at the cooling coil:
	\begin{align}\label{eq:qDotCCUB}
		\qDotCC_{\UB,k} = C_{\pw} \mDotLW_k (\TLWR_{\UB}-\TLWS_k),
	\end{align}
 where  $\TLWR_{\UB}$ is the maximum
        allowable outlet water temperature, which implicitly defines
        the capacity of the cooling coil for a given chilled water
        flowrate $\mDotLW$, and $\mDotLW_{\UB}$ is the maximum
        allowable chilled water flowrate due to cooling coil pipe
        size. The inputs need to satisfy the following additional constraints:
	\begin{align}\label{eq:CC_InputCons}
		\TLWS_{k}\leq \TLWR_{\UB},\quad 0\leq \qDotL_{k},\quad 0 < \mDotLW_k \leq \mDotLW_{\UB},
	\end{align}
	where the first two constraints prevent heating in the cooling
        coil and the last constraint prevents division by zero.
	
	The
        equations~\eqref{eq:cc-TLWR},~\eqref{eq:cc-qdotcc},~\eqref{eq:qDotCCUB},
        and~\eqref{eq:CC_InputCons}  constitute the proposed aggregate cooling
        coil model. Given the current state and inputs (command and disturbance),
        the next state $x^{\cc}_{k+1}$ can be
        computed from them.

        The model is an abstraction of the action of the local
                controller at the cooling coil that tries to maintain
                the supply air temperature at its setpoint as long as
                the chilled water stream can deliver sufficient
                cooling. When the cooling load is higher than the
                capacity of the cooling coil, the coil will only
                deliver the maximum possible heat transfer rate, i.e.,
                $\qDotCC_k = \qDotCC_{\UB,k}$. This way, the model
                incorporates heat exchanger saturation:

Multiple cooling coils can be modeled by combining multiple models of
the form above. The aggregate $\TLWR$ will be the weighted average
(weighted by flowrate) of the outlet temperatures at each cooling
coil. The model above can also be replaced with a more complex model,
such as that in ~\cite{RamanMPC_AE:2020}, as long as the same inputs, states, and outputs are included in the new model. 

	\begin{remark}
		The reason for enforcing heat transfer capacity of the
                cooling coil implicitly via the maximum return
                temperature of the ``load water'' is that, since the
                cooling coil model is an aggregate of many coils, the
                upper bound on its heat exchange capacity is the sum
                of these individual capacities, which requires
                considerable bookkeeping that is prone to error. In
                contrast, $\TLWR_{\UB}$ is fairly standard, and is
                thus easy to specify. According to the ASHRAE
                handbook~\cite{ASHRAE_Standard-90-1} this temperature
                is typically maintained at $12-15\degree$C. In the sequel, we use
                this approach for enforcing capacity of chiller and
                cooling tower as well.
	\end{remark}
	
	\subsection{Thermal Energy Storage (TES) Model}
	The thermal energy storage system considered here is a chilled
        water tank. The TES is modeled by two virtual sub-tanks: warm water
        tank (superscript ``tww'') and cold water tank (superscript
        ``twc''), as shown in Figure~\ref{fig:DetailCCWP}. This is a
        common modeling approach since a thermocline separates the
        warm water and cold water in the
        TES~\cite{ASHRAE_TESguide:2019}. Associated with each tank are
        two variables: the temperature of the waters in the sub-tanks: $\TTWW_k$ and
        $\TTWC_k$, and the ``fractions'' of water mass in them: $\STWW_k$ and
        $\STWC_k$. The total mass of water inside the TES is always
        kept constant, i.e., $\STWW_k + \STWC_k = 1$. The amount of
        increase in one sub-tank equals to the amount of decrease in
        the other sub-tank. The flow rate of water into the TES is
        $\mDotTW_k$. When $\mDotTW_k > 0$, this means that $\mDotTW_k
        t_s$ (kg) of water at a temperature $\TSW$ (supply water from
        the chillers; see Fig.~\ref{fig:DetailCCWP}) is charged into
        the cold water tank at time $k$ (rather, during the time
        interval $[(k-1)t_s, kt_s]$) and $\mDotTW_k t_s$ (kg) of water
        at temperature $\TTWW$ has been discharged from warm
        water tank during the same time. When $\mDotTW_k < 0$,
        $\mDotTW_k t_s$ (kg) of water at temperature $\TLWR$ (return
        water from the cooling coil) is
        charged into the warm water tank and $\mDotTW_k t_s$ (kg) of
        water at a temperature of $\TTWC$ has been discharged from
        cold water tank. We assume the TES is well insulated and that
        no mixing occurs between the two sub-tanks. Thus, the
        temperature in each sub-tank is only effected by the
        temperature and flow rate of water coming into and out of the
        TES. The input $u^{\tes}$ and state $x^{\tes}$ of the TES model are:
	\begin{align}
		u^{\tes} &= [\TSW, \TRW, \mDotTW]^T  \label{eq:uTES} \\ 
		x^{\tes} &= [\TTWC, \STWC, \TTWW, \STWW]^T, \label{eq:xTES}
	\end{align}
	Mass balance leads to the following equations for the water fractions:
	\begin{align} 
		\STWC_{k+1} &= S^\twc_k + \frac{\mDotTW_kt_s}{\MTES},   \label{eq:S_TWC}\\ 
		\STWW_{k+1} &= S^\tww_k - \frac{\mDotTW_kt_s}{\MTES}, \label{eq:S_TWW}
	\end{align}
where $\MTES$ (kg) is the total water mass in the TES.	For the warm water sub-tank, heat balance leads to
	\begin{align}\label{eq:TES-intermediate} 
		\MTES\big(\TTWW_{k+1}\STWW_{k+1}-\TTWW_{k}\STWW_{k}\big) = 
		\begin{cases}
			-t_s \mDotTW_k \TLWR_k, & \mDotTW_k < 0 \\
			-t_s \mDotTW_k \TTWW_k, & \mDotTW_k > 0
		\end{cases}.
	\end{align}
	By using~\eqref{eq:S_TWW}, we can express~\eqref{eq:TES-intermediate} compactly as:
	\begin{align}\label{eq:T_tww}
		T^\tww_{k+1} = T^\tww_{k} + t_s \frac{\text{min}(\dot{m}^\tw_k,0)}{\MTES S^\tww_k - t_s\dot{m}^\tw_k}\Big(T^\tww_k - \TLWR_k\Big).
	\end{align}
	A similar derivation for the cold water tank leads to:
	\begin{align}\label{eq:T_twc}
		T^\twc_{k+1} = T^\twc_{k} + t_s \frac{\text{max}(\dot{m}^\tw_k,0)}{\MTES S^\twc_k + t_s\dot{m}^\tw_k}\Big(T^{\sw}_k - T^\twc_k\Big).
	\end{align}
The following must also hold:
	\begin{equation} \label{eq:TES_InputCons}
		\left.\begin{aligned}
			&\STW_{\LB} \leq \STWC_k \leq \STW_{\UB} \\
			&\STW_{\LB} \leq \STWW_k \leq \STW_{\UB}
		\end{aligned}\right\}  \text{for all $k$}
            \end{equation}
            where $\STW_{\UB}$ and $\STW_{\LB}$ are maximum and
            minimum water mass fractions for each sub-tank.
            	The equations~\eqref{eq:S_TWC}, \eqref{eq:S_TWW},
                \eqref{eq:T_tww}, \eqref{eq:T_twc}, and
                \eqref{eq:TES_InputCons} constitute the TES
                model. Given the current state and input, one can
                compute the next state from these equations.
	
	\subsection{Chiller Model} \label{sec:ChillerModel}

\begin{figure}
		\centering
		\includegraphics[width=0.95\columnwidth]{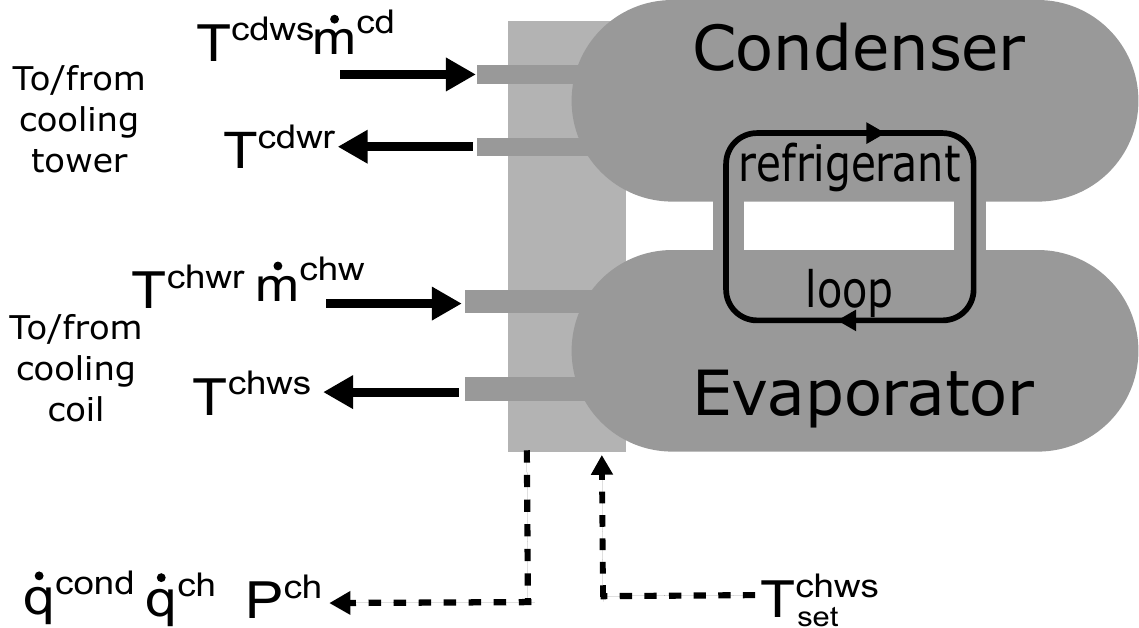}
		\caption{Schematic of a water-cooled vapor compression
                  chiller, showing variables used in the model.}
		\label{fig:chiller}
	\end{figure}

	We consider water-cooled vapor compression chillers; see
        Fig.~\ref{fig:chiller}. The refrigerant loop in the chiller transfers heat from the
        chilled water stream to the cooling water stream, which
        reduces the chilled water temperature from $\TChWR$ to
        $\TChWS$ and increases the condenser water temperature from  $\TCONDWS$ to $\TCONDWR$. 

        The proposed chiller model is modified from the
        \ElectricEIR\ chiller model
        in~\cite{DOEEnergyPlusEngRef:2022}. The \ElectricEIR\
        chiller model abstracts out the details of the refrigerant
        loop, and assumes that all of the heat absorbed by the chilled
        water at the chiller evaporator is transferred to the
        condenser water at the chiller
        condenser. Since it does not limit the condenser capacity,
        unrealistically high condenser water return temperature $\TCONDWR$
        can be predicted when cooling by the condenser water is inadequate, as demonstrated in
        Section~\ref{sec:OriginalChillerFail}. We modify the
        \ElectricEIR\ chiller model by enforcing saturation of the condenser 
 water outlet temperature at the predetermined value $\TCWRUB$, which implicitly
        defines a hard upper bound on the capacity of the condenser, i.e., $\qDotCond_{\UB}$. Because of the saturation feature,        the proposed model is called the \OurElectricEIR\ model.

	A chiller does not have exogenous inputs; the control input and state of a chiller are:
	\begin{align}
		u^\ch &= [\TChWR, \mDotCHW, \TCONDWS, \mDotCOND, \TChWSSet]^T,  \label{eq:uCh} \\
		x^{\ch} &= [\TChWS, \TCONDWR]^T, \label{eq:xCh}
	\end{align}
	where $\TChWR$ and $\mDotCHW$ are the temperature and flowrate
        of inlet chilled water at the evaporator, $\TCONDWS$ and
        $\mDotCOND$ are the temperature and flowrate of cooling
        water entering the chiller condenser, $\TChWSSet$ is the
        setpoint for the temperature $\TChWS$ of the chilled water at the
        evaporator outlet, $\TCONDWR$ is the temperature at the
        condenser outlet; $\qDotEvap$ and $\qDotCond$ are the rates of heat exchanges at the evaporator and the condenser; $\PCh$ is the power consumption of the chiller; $\TCWRUB$ is the maximum allowable condenser outlet water temperature; $\qDotEvapNoml$ is the nominal chiller (evaporator) capacity and $\PChNoml$ is the nominal chiller power consumption.
                
	The chilled water supply temperature
        $\TChWS_{k+1}$ (at evaporator) is modeled based on heat
        balance with a first order dynamic behavior: 
	\begin{align}
		\TChWS_{k+1} &= a^{\ch}\TChWS_{k} + (1- a^{\ch})(\TChWR_k - \frac{\qDotEvap_k}{C_{\pw} \cdot \mDotCHW_k}). \label{eq:TChWS}
	\end{align}
        where $a^{\ch}$ is the coefficient of a low pass filter used to model first order response between inlet and outlet temperatures at the evaporator. The cooling provided by evaporator $\qDotEvap_k$ is not
        determined yet, but we know it cannot exceed either its upper
        bound $\qDotEvap_{\UB,k}$ or the required cooling power $\qDotEvap_{\req,k}$:
	\begin{align}
	0 \leq	\qDotEvap_k &\leq \min \{\qDotEvap_{\req,k}, \ \qDotEvap_{\UB,k}\}. \label{cons:qDotEvap} 
	\end{align}
	The required cooling rate $\qDotEvap_{\req,k}$ is determined by the difference between the chilled water return temperature and the chilled water supply setpoint: 
	\begin{align} \label{eq:qDotEvapReq}
		\qDotEvap_{\req,k} &= C_{\pw} \cdot\  \mDotCHW_k \cdot\  (\TChWR_k-\TChWS_{\set}).
	\end{align}	
	The upper bound on the evaporator heat exchange rate capacity is:
	\begin{align}\label{eq:chillerCapacityDef}
		\qDotEvap_{\UB,k} &= \qDotEvapNoml \cdot\ \CapFunT_k,
	\end{align}
	where $\qDotEvap_{\noml}$ is the constant nominal capacity of the chiller (evaporator); $\CapFunT_k$ is a capacity modifier function that depends on myriad factors and is usually an empirical relationship that differs from one chiller to another. We take the empirical relationship provided for the \ElectricEIR\ chiller in~\cite{DOEEnergyPlusEngRef:2022}:
	\begin{align}\label{eq:CapFunT}
		\CapFunT_k &= \alpha_1 + \alpha_2\TChWS_{k} + \alpha_3(\TChWS_{k})^2 + \alpha_4\TCONDWS_k \nonumber\\ & + \alpha_5(\TCONDWS_k)^2 + \alpha_6 \TChWS_{k} \TCONDWS_k, 
	\end{align}
        where $\alpha_1,..., \alpha_6$'s are empirical parameters.
	For the chiller condenser, a heat balance leads to the
        following dynamic model of the condenser water return
        temperature $\TCONDWR$ with an assumed first order response
        between the input and the output (with filter coefficient $a^{\cond}$):
	\begin{align} 
		\TCONDWR_{k+1} & = a^{\cond}\TCONDWR_k + (1- a^{\cond})(\TCONDWS_k + \frac{\qDotCond_k}{C_{\pw}\cdot \mDotCOND_k}), \label{eq:TCWR}
	\end{align}
	where $\mDotCOND_k$ is the flowrate of cooling water through the condenser, and the heat exchange at the condenser $\qDotCond_k$ is:
	\begin{align} 
		\qDotCond_k &= \qDotEvap_k + \eta_1 \PCh_k, \label{eq:qDotCond}
	\end{align}
	where the second term $\eta_1\PCh_k$ is due to the waste heat produced by the electricity consumed by the chiller, mainly the compressor motors. The condenser return water temperature cannot be lower than the supply (after picking up heat at the condenser), and it cannot exceed its upper bound:
	\begin{align}
		\TCONDWS_k  &\leq \TCONDWR_{k}\leq \TCONDWR_{\UB}, \label{cons:TCWRLBUB}
	\end{align}
	In the \ElectricEIR\ model
        from~\cite{DOEEnergyPlusEngRef:2022}, the electric power
        consumed by the chiller, $\PCh_k$, is modeled by  the
        following empirical relationships:
	\begin{align}\label{eq:P_CH}               
		\PCh_k =& \PChNoml \cdot \CapFunT_k \cdot\ \EIRFunT_k \cdot\ \EIRFunPLR_k \cdot \CR_k
	\end{align}
	where $\CapFunT$ is defined in~\eqref{eq:CapFunT}, and
	\begin{align}\label{eq:EIRFunT}
		\EIRFunT_k &= \beta_1 + \beta_2\TChWS_{k} + \beta_3(\TChWS_{k})^2 + \beta_4\TCONDWS_k \nonumber \\
		&+ \beta_5(\TCONDWS_k)^2 + \beta_6 \TChWS_{k} \TCONDWS_k,
	\end{align}
        where $\beta_i$'s are empirical parameters. The  $\EIRFunPLR_k$ term is the modifier due to the part load ratio:
	\begin{align}
		\EIRFunPLR_k &= \gamma_1 + \gamma_2 \PLR_k + \gamma_3 \PLR_k^2, \label{eq:EIRFunPLR} \\
		\PLR_k &= \min\big\{\max\{ \frac{\qDotEvap_{k}}{\qDotEvap_{\UB,k}}, \PLR_{\LB}\} , \PLR_{\UB}  \big\},   \label{eq:PLRBounds}
	\end{align}
        with coefficients $\gamma_i$'s.	The last term $\CR$ in~\eqref{eq:P_CH} is the cycling ratio:
	\begin{align} \label{eq:CR}
		\CR_k \eqdef \min\{\frac{\PLR_k}{\PLR_{\LB}}, \ 1.0\}
	\end{align}
	which is the modifier to account for chiller cycling on and
        off under very low load condition, where $\PLR_{\LB}$ and $\PLR_{\UB}$ are lower and upper bounds for the part load ratio.
	
	Note that given the current input and state
        $u_k^{\ch},x_k^{\ch}$, the next state $x^{\ch}_{k+1}$, namely
        $\TChWS_{k+1}$ and $\TCWR_{k+1}$, cannot be uniquely
        determined from these relationships since $\qDotEvap_k$ is not
        yet known. The intermediate variable $\qDotEvap_k$ and the
        next state $x_{k+1}^{\ch}$ will be determined together as
        follows. The low-level controllers will try to run the chiller
        to cool down the chilled water from $\TChWR_k$ to the setpoint
        $\TChWSSet$ unless either the evaporator reaches its capacity ($\qDotEvap_k =
        \qDotEvap_{\UB,k}$) or the condenser reaches its capacity
        ($\TCONDWR_{k+1}=\TCONDWRUB$). The actions of these low-level
        controllers - along with the associated physics that enforces
        heat exchanger capacities - are abstracted into the following
        optimization problem: 
	\begin{align}\label{eq:ChillerModel}
		\begin{split}
			\min_{\qDotEvap_k} & (\TChWS_{\set,k} - \TChWS_{k+1})^2 \\
			\text{ s.t. } &\eqref{eq:TChWS} - \eqref{eq:CR}
		\end{split}
	\end{align}
	In summary,~\eqref{eq:ChillerModel} - a nonlinear program
        (NLP) - is the proposed chiller
        model as its solution determines the next state
        $x^{\ch}_{k+1} = (\TChWS_{k+1},\TCONDWR_{k+1})$ from $u_k^{\ch},x_k^{\ch}$.

	        Additionally, the problem data must satisfy
	\begin{align}
		\TChWS_{\LB}&\leq \TChWS_{\set,k}\leq \TChWR_k,  \label{cons:TChWSSet<TChWR}\\
		\TCWS_k &\leq \TCONDWR_{\UB},  \label{cons:TCWS<TCWRUB}
	\end{align}
	where \eqref{cons:TChWSSet<TChWR} ensure $\qDotEvap_{\req,k}
        \geq 0$ and \eqref{cons:TCWS<TCWRUB} ensures $\qDotCond_k\geq
        0$. If the inputs do not satisfy one of the
        constraints~\eqref{cons:TChWSSet<TChWR}-\eqref{cons:TCWS<TCWRUB},
        the chiller's refrigerant loop will be considered inoperative, and chilled water
        and condenser water will pass through the chiller with no heat exchange: $\TChWS_{k+1} = \TChWR_k$ and $\TCONDWR_{k+1} = \TCONDWS_k$.

        The electrical power consumed by the pumps that circulate the water streams in the chillers is modeled using a  black-box model proposed in~\cite{RisbeckMixedintegerEnB:2017}:
\begin{align}
	\PChWP_k &=
                   a_1\ln(1+a_2\dot{m}_k^{\chw})+a_3\dot{m}_k^{\chw}+a_4, \label{eq:PChWP}
\end{align}
where $a_i$ are empirical parameters.

        \begin{remark}[Difference
          between the proposed chiller model \OurElectricEIR\ and the
          original \ElectricEIR\ model]
The upper bound $\TCWR_{\UB}$ in
        equation~\eqref{cons:TCWRLBUB} implicitly enforces an upper
        bound on the heat exchange rate $\qDotCond_k$ at the condenser
        in equation~\eqref{eq:TCWR}: the condenser water can absorb
        heat up to the predetermined maximum possible, if needed, but  not
        more. How much it will in fact absorb under part load
        conditions is computed by solving the
        NLP~\eqref{eq:ChillerModel}. This feature is not present in the \OurElectricEIR\ model, which led to the behavior discussed in Section~\ref{sec:OriginalChillerFail}.
        \end{remark}

	\paragraph{Numerical verification}
For numerical verification, we use a chiller of type
\texttt{Electric:EIR:Chiller Carrier 19XR
  823kW/6.28COP/Vanes}~\cite{DOEEnergyPlusEngRef:2022}, with 823
kW-thermal rated capacity. Since the main new feature in the proposed
model is the optimization-based formulation to predict capacity
saturation and thereby avoid the kind of failures of existing models
described in Section~\ref{sec:OriginalChillerFail}, here we only
verify this feature.         

\begin{figure}[ht]
		\centering
		\subfigure[junk][
		\label{fig:T_CWR_Comparison}]
		{
                \includegraphics[width=0.9\columnwidth]{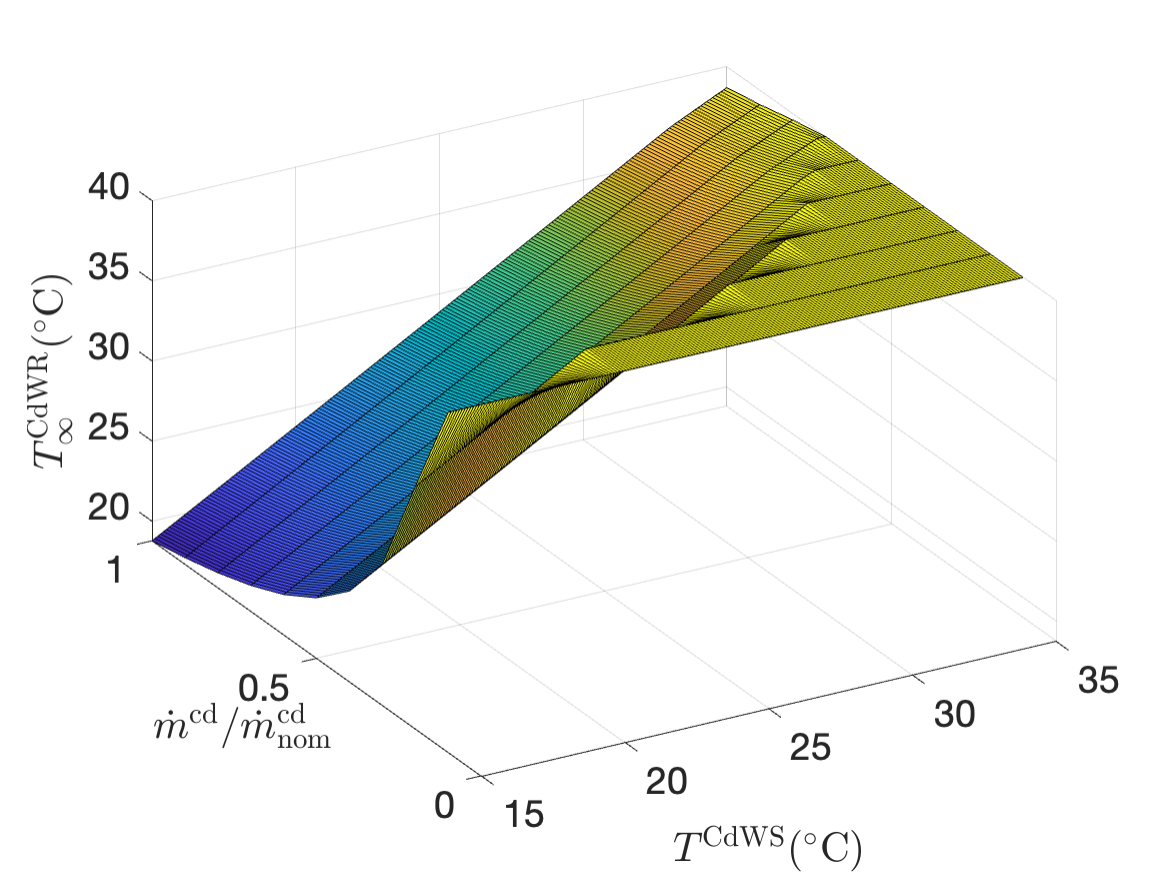}
		}
\subfigure[junk][\label{fig:sat_T_ChWS_Comparison}]
{
  \includegraphics[width=0.9\columnwidth]{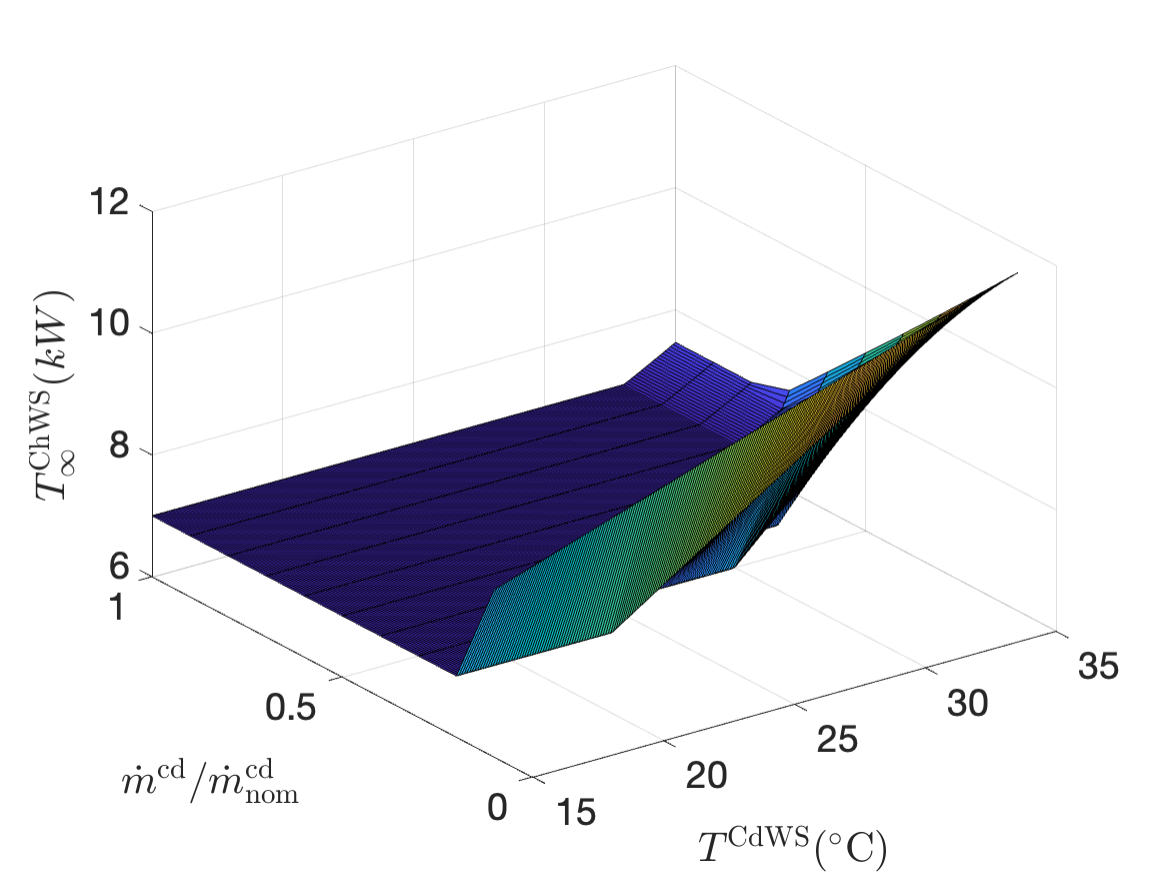}
  }
		\caption{Prediction of the proposed
                  \OurElectricEIR\ chiller model: (top) steady state condenser outlet temperature and
                  (bottom) evaporator outlet temperature 
                  (vertical axis) as a function of two inputs
                  (horizontal axes), while other inputs are held
                  constant ($\TChWR = 12\degree$C, $\mDotCHW = 31.86$ kg/s). The value of $\mDotCOND_{\noml}$ is
                  47.44 kg/s.}
	\end{figure}

Figures~\ref{fig:T_CWR_Comparison} shows prediction of the steady state condenser water return temperature
a function of condenser water supply temperature and flow rate, while the other inputs are fixed.
We see that when the condenser water flow rate becomes sufficiently small, the
proposed model predicts that the return temperature
saturates at its upper bound, $40\degree$C. Meaning, the heat exchanger in the condenser
has reached its maximum capacity and cannot remove any more heat from
the refrigerant loop. As a result, the chilled water supply temperature on the evaporator side becomes
quite warm; see Figure~\ref{fig:sat_T_ChWS_Comparison}. In contrast,
the original model does not have this saturation and thus predicts
incorrectly that cooling water return temperature may reach
90$\degree$C; cf.~Fig.~\ref{fig:CoolingWaterBlowup-combined}. These
numerical tests thus verify that the proposed chiller model indeed
extends the region of validity over the existing \ElectricEIR\ model.

\begin{figure}[ht]
          \centering
          \subfigure[At the extreme values of the inputs, the cooling water return temperature saturates to its upper bound,  $40\degree$C, since the chiller condenser has reached its heat exchange capacity][
		\label{fig:ChModelVerif_TCWR}]
		{
                  \includegraphics[width=0.9\columnwidth]{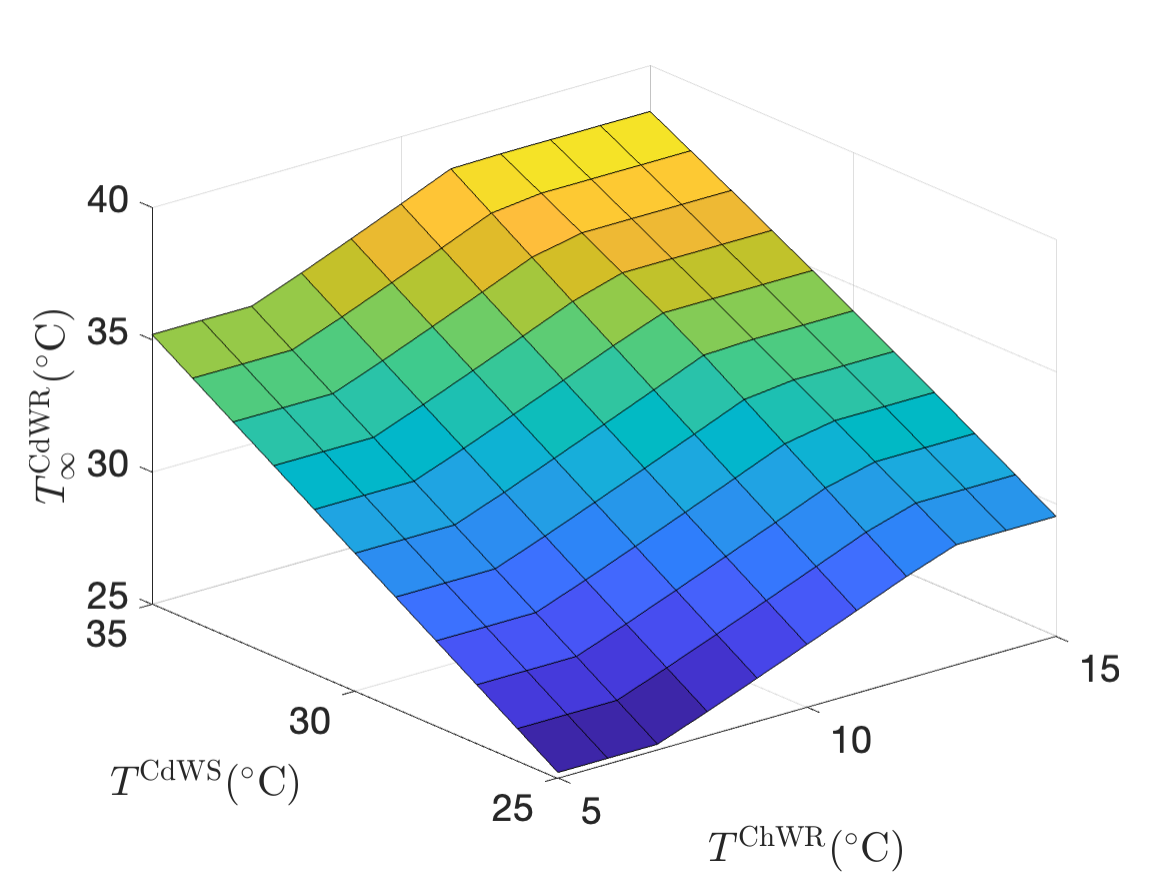}
                }
                \subfigure[The chiller evaporator is able to keep the chilled water supply at the setpoint, $7\degree$C, only for certain values of the inputs. Beyond that, the evaporator has reached its heat exchange capacity and thus the chilled water supply temperature is warmer.][
                \label{fig:ChModelVerif_TChWS}]
                {
                  \includegraphics[width=0.9\columnwidth]{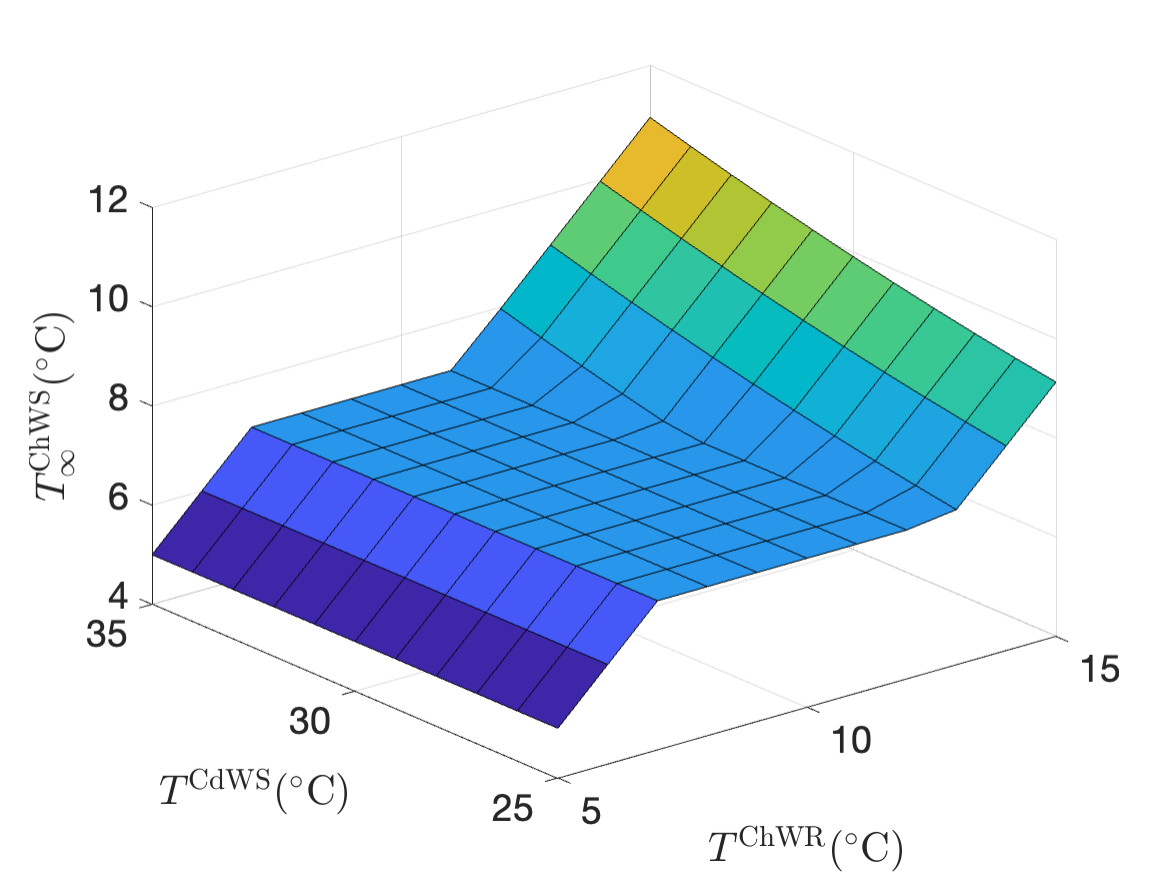}
                }	
		\caption{The proposed chiller model \OurElectricEIR's prediction (top) condenser outlet temperature and
                  (bottom) evaporator outlet temperature at steady state
                  (vertical axis) as a function of two inputs 
                  (horizontal axes), while other inputs are held constant ($\mDotCHW = 31.86$ kg/s, $\mDotCOND= 47.44$ kg/s.)}
		\label{}
	\end{figure}

        Figures~\ref{fig:ChModelVerif_TChWS} and~\ref{fig:ChModelVerif_TCWR}
show the steady-state prediction of the two states variables of the
proposed chiller model, both showing heat exchanger capacity
saturation at extreme values of inputs. 
	
	\subsection{Cooling Tower Model}
\begin{figure}
		\centering
		\includegraphics[width=0.8\columnwidth]{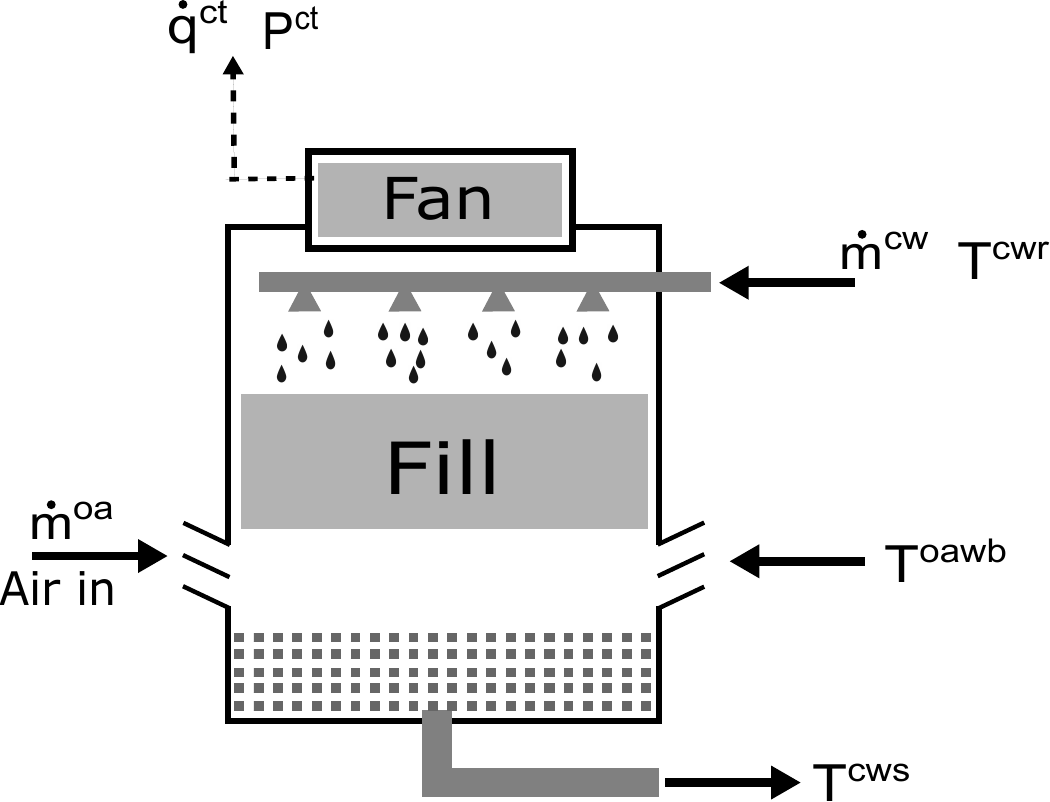}
		\caption{Cooling tower flows and their
                  related variables. Inlet and outlet water flow rates
                are assumed same.}
		\label{fig:CT}
	\end{figure}

%	A cooling tower sprays the (warm) cooling water returned from the chiller by passing it together with  ambient air through a fill; see Figure~\ref{fig:CT}. During this process, a small amount of water spray will evaporate into the air, removing heat from the cooling water, thereby cooling the water. The water loss due to its evaporation is replenished by fresh water, thus we assume the supply water flow rate equals to the return water flow rate at the cooling tower. A fan or a set of fans is used to control the ambient airflow at the cooling tower.
	We consider variable speed evaporative cooling towers in this
        paper; see Fig.~\ref{fig:CT}.
	The inputs and state of the cooling tower model are:
	\begin{align}
		u^\ct &= [\TCWR, \mDotCW, \mDotOA]^T, \label{eq:uCT} \\
		w^\ct &= \TOAWB,\\		
		x^{\ct} &= \TCWS,
	\end{align}
        where $\TCWS, \TCWR$ are the temperatures of the cooling water supply (from the cooling tower to the chiller condensers) and cooling water return (to the cooling tower); see Fig.~\ref{fig:DetailCCWP}.
        As in the other equipment, the actual heat exchange in the
        cooling tower will be the smaller of the heat exchange desired
        (decided by a setpoint to a local controller that manipulates
        the fan speed) and the capacity of the heat exchanger which is
        a function of the OA wet bulb and air flow rate through the tower.

	The proposed cooling tower model is modified from the
        variable speed evaporative cooling tower model in EnergyPlus~\cite{DOEEnergyPlusEngRef:2022} with the so-called
        ``YorkCalc'' correlations. In the sequel, we will call the
        unmodified model from EnergyPlus the \YorkCalc\ model, and proposed model the
        \YorkCalcOurs\ model, meaning, ``with saturation''. A few
        intermediate variables are needed to describe the model. Define the \emph{scope} temperature $T^{\sco}$ as the difference between the hot cooling water return temperature and the OA wt bulb temperature:
	\begin{align}                                                                                                             
		T^{\sco}_k &:= T^{\cwr}_k - T^{\oawb}_k,\label{eq:Tsco}
	\end{align}
	which is a function purely on the inlet conditions ($u^{\ct}$ and $w^{\ct}$). Define also the \emph{range} temperature $\TRan$, and \emph{ approach} temperature $\TAppState$ of the cooling tower as:
	\begin{align}                                                                                                             
		\TRanState_k &:= T^{\cwr}_k - T^{\cws}_{k},\label{eq:Tran} \\
		\TAppState_k &:= T^{\cws}_{k} - T^{\oawb}_k.\label{eq:Tapp} 
	\end{align}
	It follows from~\eqref{eq:Tsco}-\eqref{eq:Tapp} that 
	\begin{align}
		\TRan_k + \TApp_k = T^{\sco}_k. \label{eq:CTBalance}
	\end{align}
	Also, $\TRanState \in [\TRan_{\LB}, \TRan_{\UB}]$, $\TAppState \in
        [\TApp_{\LB}, \TApp_{\UB}]$ where the lower and upper bounds
        are known constants for a particular cooling tower design. There are three possible scenarios:
	\paragraph{Case 1} $\TCWR_k - \TOAWB_k < \TRan_{\LB}+\TApp_{\LB}$: The temperature difference between OA (wet bulb) and cooling water return is not adequate to lose heat to the atmosphere. Meaning,  water vapor cannot evaporate, and no cooling can be provided to the cooling water. Therefore, $\TCWS_{k+1} = \TCWR_k$.
	\paragraph{Case 2} $\TCWR_k - \TOAWB_k>  \TRan_{\UB}+\TApp_{\UB}$: This is the other extreme, in which the maximum possible range and approach still cannot satisfy~\eqref{eq:CTBalance}. In this case, the range is saturated at its maximum: $\TCWS_{k+1} = \TCWR_k - \TRan_{\UB}$. 
	\paragraph{Case 3} $\TCWR_k - \TOAWB_k \in [\TRan_{\LB}+\TApp_{\LB}, \TRan_{\UB}+\TApp_{\UB}]$:
	In the YorkCalc model, the following empirical relationship is used for computing the approach for a given range~\cite{DOEEnergyPlusEngRef:2022}:
	\begin{align}\label{eq:func_Tapp}
		&\widehat{T}^{\app}(\TRan, u^{\ct},w^{\ct}) \nonumber \\
		&= \sum_{i=0, j=0,\ell=0}^{2} c_{(i+1)\cdot (j+1)\cdot (k+1)}(\TOAWB)^i (\TRan)^j LGR^\ell
	\end{align}
	where  $LGR$ is the liquid gas ratio :
	\begin{align}
		LGR_k &= \min \{\frac{FR^{\wat}_k}{FR^{\air}_k}, LGR_{\UB} \}, 
	\end{align}
	where $FR^{\air}$ and $FR^{\wat}$ are the fractions of air and water flowrates:
	\begin{align}
		FR^{\wat}_k &= \frac{\mDotCW_k}{\mDotCWNoml} , &		FR^{\air}_k &= \frac{\mDotOA_k}{\mDotOANoml} 
	\end{align}
	Our proposal is to compute range and approach temperatures by solving the following optimization problem:
	\begin{align}\label{eq:SolveT_ran}
          \begin{split}
        	& \min_{\TRan_k, \TApp_k}  (\TAppState_k - \widehat{T}^{\app}(\TRanState_k,u^{\ct},w^{\ct}))^2 \\
		\text{ s.t. } & \ \TRan_{\LB} \leq  \TRanState_k \leq \TRan_{\UB}, \\
		& \eqref{eq:CTBalance}: \TRanState_k + \TAppState_k = T^{\sco}_k , \\
		& \TApp_{\LB} \leq \TAppState_k \leq \TApp_{\UB}.  
          \end{split}
        \end{align}
        That is, we use the empirical relation~\eqref{eq:func_Tapp} as an approximation. Once the range temperature $\TRanState_k$ is computed by solving the
        NLP~\eqref{eq:SolveT_ran}, the cooling water supply
        temperature $\TCWS_{k+1}$ is determined from:
        \begin{align}
          \label{eq:1}
          \TCWS_{k+1} = a^{\ct}\TCWS_{k} + (1-a^{\ct})(\TCWR_k - \TRanState_k)
        \end{align}
where $a^{\ct}$ is a low pass filter coefficient to model a first order
response between inlet and outlet temperatures in the heat
exchanger. The heat exchanged from cooling water to the ambient in the
cooling tower is:
	\begin{align} \label{eq:qCT}
		\qDotCT_k &= C_{\pw} \mDotCW_k (\TCWR_{k} -\TCWS_{k})
	\end{align}
	The electrical power consumption by the fan motors at the
        cooling tower is modeled using cubic fan law~\cite{DOEEnergyPlusEngRef:2022}:
	\begin{align} \label{eq:PCT}
		\PCT_k &= \PCTNoml \cdot (FR^{\air}_k)^3.
	\end{align}
        The electrical power consumption by the water pumps that circulate the cooling water is modeled using a  black-box model proposed in~\cite{RisbeckMixedintegerEnB:2017}:
\begin{align}
	\PCWP_k &= \gamma_1 \ln(1+\gamma_2\dot{m}_k^{\cw})+\gamma_3\dot{m}^{\cw}_k+\gamma_4.\label{eq:PCWP} 
\end{align}
where $\gamma_i$'s are empirical parameters.

	\begin{remark}[Difference between \YorkCalc\ model in EnergyPlus and
                MBL and the proposed \YorkCalcOurs\ model] The
                NLP~\eqref{eq:SolveT_ran} ensures that when the prediction from the
                YorkCalc correlations are infeasible (such as when
                $\widehat{T}^{\app}$ is negative), range and approach
                temperatures are still within pre-specified
                bounds. % Solving the NLP still requires iterative
                % computations, as the  \YorkCalc\ model does. But the
                % NLP solver handles them.
\end{remark}

\begin{figure}[ht]
	\centering
	\includegraphics[width=0.9\columnwidth]{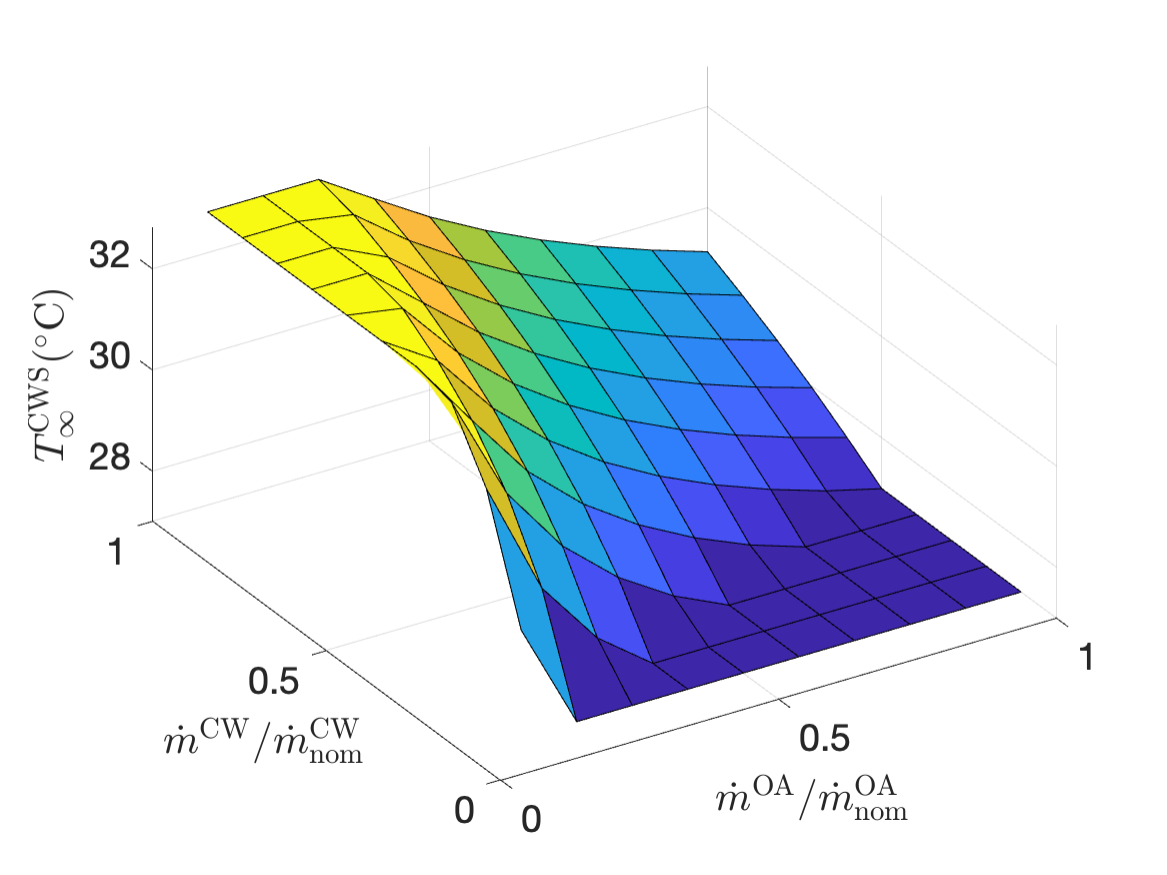}
	\caption{Cooling tower model prediction: $\TCWS_\infty$ as a function of the two mass
          flow rates while other inputs are held constant.}
	\label{fig:CTmodelVerif_T_CWS}
\end{figure}

\paragraph{Numerical Verification}
A tower of YorkCalc type with 47.44 kg/s nominal chiller
condenser water flow rate is chosen. The other nominal values are return water
temperature: $35\degree$C, supply water temperature $29.44\degree$C, and OA
wet bulb temperature:
$25.5\degree$C~\cite[Sec. 40.17]{ASHRAE_handbook_fund:16}. Thus, the
nominal approach and range are $3.88\degree$C and $5.55\degree$C. We take $\TRan_{\LB} = 2.2\degree$C, meaning
no cooling takes place if the range temperature is below $2.2\degree$C. Figure~\ref{fig:CTmodelVerif_T_CWS} show the prediction of the steady state $\TCWS$ for varying
values of the air and water flow rate fractions, while $\TOAWB_k,\TCWR_k$ are
held fixed at $25.5\degree$C and $35\degree$C. We see from the figure that when the OA flow rate is
really low and the cooling water flow rate is high, the proposed
model correctly predicts that the cooling tower would not be able to
provide sufficient cooling and that the return water from
the cooling tower would be at the same temperate as that of the water supplied to it ($35\degree$C).

% \subsection{Water pump model}\label{sec:PumpModel}
% Water pumps cycle the chilled water and cooling water in the \plant. The black-box model proposed in~\cite{RisbeckMixedintegerEnB:2017} is used to model the electrical power consumption of water pumps at the chillers and cooling towers:
% \begin{align}
% 	\PChWP_k &= \alpha_1\ln(1+\alpha_2\dot{m}_k^{\chw})+\alpha_3\dot{m}_k^{\chw}+\alpha_4, \label{eq:PChWP} \\
% 	\PCWP_k &= \gamma_1 \ln(1+\gamma_2\dot{m}_k^{\cw})+\gamma_3\dot{m}^{\cw}_k+\gamma_4.\label{eq:PCWP} 
% \end{align}
% where $\alpha_i,\gamma_i$ are empirical parameters.

\section{Central Chilled Water Plant Model}\label{sec:FullModel}
The widely deployed primary-secondary pumping strategy is considered
in our \plant\ model~\cite{ASHRAE_Handbook_HVACApplication}. Two
subsystems --- chilled water loop and cooling water loop --- are first
constructed by interconnecting equipments in these water loops, whose
models were described in Section~\ref{sec:equip}. The interconnection
of the two water loops leads to the full \plant\ model.

Since there are multiple chillers and cooling towers in the
        \plant, the letter $i$ is used as the index for each chiller
        and cooling tower. On/off status of an individual equipment is denoted by the symbol
$\mathbbm{1}$. For instance, $\mathbbm{1}^{\ch,i} =1$ if chiller $i$ is ``on'' and 0 if
``off; similarly $\mathbbm{1}^{\ct,i}$ is used to denote on/off status
of the $i$-th cooling tower. A variable without an index is the sum
        of those with indices. For example, $\mDotCHWOne$ is the
        flowrate of chilled water through chiller 1, whereas
        $\mDotCHW$ is the total chilled water flowrate through all the ``on''
        chillers.

\subsection{Chilled water loop} \label{sec:CHWLoopModel}
The chilled water loop (see Fig.~\ref{fig:DetailCCWP} consists of
four types of equipment: an aggregate cooling coil, a TES, chillers, and
chilled water pumps. The exogenous inputs and states of all the constituent component models form those of the chilled water loop. However, due to the interconnected nature of the system the output of an equipment, or the combination of the outputs of multiple equipments, can become the input to another equipment in this loop. So some of the inputs to component models become internal variables - that we retain as outputs - but do not count as external inputs/control commands. The commanded inputs to the chilled water loop are 
\begin{align}  \label{eq:u-chw-loop}
u^{\chw}  = [\mDotLW, \mDotTW, \TChWSSet, \mDot^{\chws,i}, \mathbbm{1}^{\ch,i},  \mDot^{\cond,i}], i=1,\dots,\nCh.
\end{align}
To fully describe the interconnected \plant\ model the following variables are needed that are considered as outputs since they are functions of the states and inputs:
 \begin{align}
     y^{\chw} &= [\TSW, \TRW, \TChWR, \TChWSav, \mDotSW, \mDotBP, \mDotCHW \nonumber ,\\ &\mDotCOND, \qDotCC, \qDotEvap, \qDotCond, \PCh, \PChWP ]. \label{eq:yCHW}
 \end{align}
To explain the components of $y^{\chw}$, first note that the flowrates, heat exchange rates, and power consumptions are the sum of corresponding terms from each individual equipment. Using the terminology described at the beginning of the section, we get
\begin{align}\label{eq:4}
  	\qDotEvap_{k} &= \sum_{i=1}^{\nCh} \qDotEvapi_{k}, &\qDotCond_{k} = \sum_{i=1}^{\nCh} \qDotCondi_k, \\
	\PCh_{k} &= \sum_{i=1}^{\nCh} \PChi_k,    &\PChWP_{k} = \sum_{i=1}^{\nCh} \PChWPi_k,
\end{align}
where $\qDotEvapi_{k}, \qDotCondi_{k}, \PChi_{k}, \PChWPi_k$ are defined
in~\eqref{eq:ChillerModel},\eqref{eq:qDotCond}, \eqref{eq:P_CH}, and \eqref{eq:PChWP}. 

In a primary-secondary pumping system, the water flowrate through each
``on'' chiller is maintained at its nominal value,
$\mDotCHWi_{\noml}$. As a result, the total chilled water flowrate
($\mDotCHW_k$) may be more than what is required by loads and TES,
which is the \emph{supply water} flowrate $\mDotSW_k$. See
the points marked ``$\sw$'' and ``$\rw$''  for ``supply water'' and
``return water'' in Fig.~\ref{fig:DetailCCWP}. A
one-way bypass valve thus routes the redundant chilled water
($\mDotBP_k$) from chiller outlet directly to the chiller inlet. We
thus obtain the following
\begin{align}
  	\mDotCHW_k &= \sum_{i=1}^{\nCh} \mathbbm{1}^{\ch,i}_k \mDotCHWi_{\noml},  &    \mDotBP_k = \mDotCHW_k -\mDotSW,   \\
\mDotSW_k  & = \mDotLW_k + \mDotTW_k, &  \mDot^{\rw} = \mDotSW. 
\end{align}
The load water supplied to the cooling coil could be the mixture of chilled
waters from the chillers or chilled water from the TES or a mixture of the two, depending
on the commands to valve ``A'' in Fig.~\ref{fig:DetailCCWP}. The return water could be the 
water leaving the cooling coil or the TES or a mixture of the two,
depending on the commands to valve ``B'' in Fig.~\ref{fig:DetailCCWP}. Chilled water return is a
mixture of return water and bypass water. The above relationships are summarized below, where $\sum_{i=1}^{\nCh}\mathbbm{1}^{\ch,i} \neq 0$ means at least one chiller is on, and $\sum_{i=1}^{\nCh}\mathbbm{1}^{\ch,i} = 0$ means
all chillers are off:
\begin{align}
	\TLWS_{k}  &= 
	\begin{cases} 
		\TSW_k + \frac{\text{min}(\mDotTW_k, 0)}{\mDotLW_k}\Big(\TSW_k - \TTWC_k\Big), &\sum_{i=1}^{\nCh}\mathbbm{1}^{\ch,i} \neq 0,  \label{eq:TLWS}  \\
		\TTWC_k, & \sum_{i=1}^{\nCh}\mathbbm{1}^{\ch,i} = 0,
	\end{cases}\\
	\TRW_{k}  &= 
	\begin{cases} 
		\TLWR_k + \frac{\text{max}(\mDotTW_k, 0)}{\mDotSW_k}\Big(\TTWW_k - \TLWR_k\Big), &\sum_{i=1}^{\nCh}\mathbbm{1}^{\ch,i} \neq 0, \label{eq:TRW}\\
		\TTWW_k, &\sum_{i=1}^{\nCh}\mathbbm{1}^{\ch,i} = 0,
	\end{cases}\\
	\TChWRav_k  &= 
	\begin{cases} 
		\TRW_k + \frac{\mDotBP_k}{\mDotCHW_k}( \TChWSav_k-T^{\rw}_k ), &\sum_{i=1}^{\nCh}\mathbbm{1}^{\ch,i} \neq 0 \label{eq:TChWRav}  \\
		\TRW_k,  &\sum_{i=1}^{\nCh}\mathbbm{1}^{\ch,i} = 0,
              \end{cases}
\end{align}
where the \emph{plantwide} chilled water supply
temperature, denoted by $\TChWSav$, is the temperature of the chilled water streams from all chillers after mixing. It is a weighted average of
the chilled water supply temperatures of all active chillers:
\begin{align}
  \label{eq: TChWSav}
  	\TChWSav_{k} &= 
	\begin{cases} 
		\frac{1}{\mDotCHW_k} \sum_{i=1}^{\nCh}
                \mathbbm{1}^{\ch,i} \TChWSi_{k}\mDotCHWi_k,
                &\sum_{i=1}^{\nCh}\mathbbm{1}^{\ch,i} \neq 0, \\
		\TChWR_k, &\sum_{i=1}^{\nCh}\mathbbm{1}^{\ch,i} = 0,
              \end{cases}
\end{align}
From  Fig.~\ref{fig:DetailCCWP}, we see that the chilled water return temperature entering each chiller's evaporator is the \emph{plantwide} chilled water return temperature:
\begin{align}
  \label{eq:TChWRi}
  \TChWRi & = \TChWRav, & i=1,\dots,\nCh.
\end{align}
Similarly,
\begin{align}\label{eq:TCONDWSi-TSW}
  \TCONDWSi &= \TCWSav  & i=1,\dots,\nCh\\
  \TSW_k &= \TChWSav_k &
\end{align}
where the plantwide cooling water supply temperature $\TCWSav$ is an output of the cooling water loop, and will be formally defined later in~\eqref{eq:TCWSav}. The flow rate of cooling water returning to the cooling water loop is the sum of the
condenser water flow rates in the chillers, and cooling water return temperature $\TCWRav$ is a weighted average of
the cooling water (condenser outlet) temperatures of all active
chillers:
\begin{align}
  \mDotCOND_k  & = \sum_{i=1}^{\nCh} \mDotCONDi_k \mathbbm{1}^{\ch,i}_k = \mDotCW_k  = \sum_{j=1}^{\nCT} \mDot^{\cw,j}_k \mathbbm{1}^{\ct,j}_k , \label{eq:mDotCWTotal}\\
  	\TCWRav_{k} &= 
	\begin{cases} 
		\frac{1}{\mDotCW_k} \sum_{i=1}^{\nCT}\mathbbm{1}^{\ch,i} \TCONDWRi_{k}\mDotCONDi_k, \quad   &\sum_{i=1}^{\nCh}\mathbbm{1}^{\ch,i} \neq 0,  \\
		T^{\cwr,\noml}_{k}, \quad   &\sum_{i=1}^{\nCh}\mathbbm{1}^{\ch,i} = 0.
	\end{cases} \label{eq:TCWRav}
\end{align}
Note that~\eqref{eq:mDotCWTotal} is due to mass conservation at the interconnection between the chilled water loop and the cooling water loop (see Fig.~\ref{fig:DetailCCWP}), which will be described next.

\subsection{Cooling Water Loop}\label{sec:CWLoopModel}
The cooling water loop consists of cooling towers and cooling water
pumps; see Fig.~\ref{fig:DetailCCWP}. The exogenous inputs and states of the cooling water loop are: $  w^{\cw} = [\TOAWB ]$, $  x^{\cw} = [x^{\ct,i}]$ where $i = 1,\dots,\nCT$. The control inputs of the cooling water loop are
\begin{align}
  \label{eq:3}
    u^{\cw}  =  [\mathbbm{1}^{\ct,j},\mDot^{\cw,j},\mDot^{\oa,j}], j = 1,\dots,\nCT.
\end{align}
As in the case of the chilled water loop, it has a few additional outputs:
\begin{align}
  y^{\cw} &= [\mDotCW,\ \TCWSav,\ \qDotCT,\ \PCT,\ \PCWP], \label{eq:yCW}
\end{align}
where
\begin{align}
  \TCWRi_k & = \TCWRav_k, i = 1,\dots,\nCT & \\
	\mDotCW_k &= \sum_{i=1}^{\nCT} \mDotCWi_k,\quad  \qDotCT_k = \sum_{i=1}^{\nCT} \qDotCTi_k, \\
	\PCT_{k}  & = \sum_{i=1}^{\nCT} \PCTi_k,  \quad  \PCWP_{k} = \sum_{i=1}^{\nCT} \PCWPi_k
\end{align}
where $\TCWRav_k, \qDotCTi_k,  \PCTi_{k}, \PCWPi_k$ are defined
in~\eqref{eq:TCWRav}, \eqref{eq:qCT}, \eqref{eq:PCT}, \eqref{eq:PCWP}, and
\begin{align}\label{eq:TCWSav}
	\TCWSav_{k} &= 
	\begin{cases} 
		\frac{1}{\mDotCW_k} \sum_{i=1}^{\nCT}\mathbbm{1}^{\ct,i}\TCWSi_{k}\mDotCWi_k, \quad &\sum_{i=1}^{\nCT}\mathbbm{1}^{\ct,i}_k \neq 0,  \\
		\TCWRav_{k},\quad  &\sum_{i=1}^{\nCT}\mathbbm{1}^{\ct,i}_k = 0,
              \end{cases}
\end{align}

\subsection{Interconnecting Water Loops to Obtain \plant\ Model } \label{sec:FullCCWPModel}
Connecting chilled water loop~\ref{sec:CHWLoopModel} and cooling water
loop~\ref{sec:CWLoopModel} together with state propagation equations (i.e., dynamic models) of the component models completes the complete model of the \plant. For
the complete \plant\ model, the plant-wide states $x^p$, inputs $u^p$, and
disturbances $w^p$, along with additional outputs, are: 
\begin{align}
	x^p &\eqdef [x^{\cc},\ x^{\tes},x^{\ch,i},x^{\ct,j}]^T \in \R^{4+2\nCh+\nCT}  \label{eq:PlantStates} \\
  u^p &\eqdef [u^{\chw},\ u^{\cw}]^T \in \mathbb{R}^{3\nCh+3\nCT+3},  \notag \\
	w^p &\eqdef [w^{\cc}, \ w^{\ct}]^T \in \mathbb{R}^{2}. \label{eq:PlantDists}  \\
  y^p &\eqdef [y^{\chw},\ y^{\cw}, \Ptot]^T \in \mathbb{R}^{18} ,\label{eq:PlantOutputs}
\end{align}
%For completeness, the control command is $u = [\mDotLW, \mDotTW, \TChWSSet, \mDot^{\chws,i}, \mathbbm{1}^{\ch,i},  \mDot^{\cond,j}, \mathbbm{1}^{\ct,j},  \mDot^{\cw,j}, \mDot^{\oa,j}] $, for $i=1,\dots,\nCh$, $j=1,\dots,\nCT$.
All variables are defined in Section~\ref{sec:CHWLoopModel} and \ref{sec:CWLoopModel} except the total electric power consumption $\Ptot$:
\begin{align}
	\Ptot_k= \PCh_k + \PCT_k + \PChWP_k + \PCWP_k. \label{eq:P_tot}
\end{align}
Together, \eqref{eq:uCh}-\eqref{eq:P_tot} describe the proposed \plant\ model:
\begin{align} \label{eq:OverallPlantModel}
	x^p_{k+1} = f(x^p_{k}, u^p_{k}, w^p_{k}), y^p_k = h(x_k,u_k,w_k)
\end{align}
A closed loop control system with a \plant\ and a supervisory
controller that decides the plant control command $u_k^p$ may have
additional exogenous input than that for the plant,  such as a forecast of
electricity price. %Certain local states and outputs of equipment are
                   %not explicitly incorporated into the plant
                   %variables, such as $\TChWSi$, $\qDotEvapi$, and $\PCWPi$. 

\subsection{Software implementation}
A \matlab\ implementation of the \plant\ model is released under the MIT
license and is available in \url{https://gitlab.com/pbarooah/ccwp_model}. The chiller model~\eqref{eq:ChillerModel} and the cooling tower
model~\eqref{eq:SolveT_ran} are non linear programs (NLPs). There are many powerful open source tools for posing and solving NLPs efficiently. In our Matlab implementation we use CasADi \cite{Andersson2019} and IPOPT \cite{wacbie:2006} to pose and solve~\eqref{eq:ChillerModel},\eqref{eq:SolveT_ran}. The user needs to download and install CasADi, which
is freely available under LGPL-3.0 license. All the parameters of the \plant\ are defined in the function \texttt{func\_defPlantParams\_2Ch2CT} in the codebase. 

                During simulation, the control inputs $u_k^p$ can be computed either by a
                supervisory controller or specified a-priori. In
                either case, not all possible values of the input
                vector $u^p_k$ are feasible. For instance, if a
                chiller is ``on'' then there must be a non-zero
                condenser water flow rate in that chiller, meaning
                $\mathbbm{1}^{\ch,1}_k=1$ and $\mDotCONDOne_k =0$ are
                not allowed. There are several such feasibility checks
                that need to be done; e.g., \eqref{eq:CC_InputCons},\eqref{eq:mDotCWTotal}. A function
                \texttt{func\_CheckInputs} in the codebase checks if
                $u_k$ passes these feasibility tests, and if not, stops the simulation.

The codebase currently has the model of a \plant\ that consists of an
aggregate load, a TES, two distinct chillers, two distinct cooling towers, and corresponding chilled water and cooling
water pumps. The parameters of \texttt{ElectricEIRChiller Carrier 19XR
  823kW/6.28COP/Vanes} model in MBL, which has a cooling capacity
0.823 MW, are used for Chiller 1. The parameters of
\texttt{ElectricEIRChiller York YT 1368kW/7.35COP/VSD} model in MBL,
which has a cooling capacity 2.1 MW, is used for Chiller 2. Both
cooling towers are of YorkCalc type, but their nominal air flowrate
and water flowrates are changed from the default to be consistent with
the chiller condenser flowrates. The design cooling capacity of the two
cooling towers combined is 2859.8 kW. The nominal power consumption of
cooling towers are modified according to MBL
references~\cite{ModelicaBldgLibrary:2014}. Some of the parameters of the \plant\ are mentioned in Table~\ref{tab:simParam}. 

The state propagation function \texttt{func\_f\_CCWP\_2Ch2CT} returns the
next state $x^p_{k+1}$ when provided $x^p_k,u^p_k,w_k^p$ and plant parameters as arguments. For
simulating a \plant\ with a different number and type of equipment, a
new state propagation function has to be written.

\begin{table}
	\centering
	\caption{A few parameters of the \plant\ model simulated.}
	\setlength{\arrayrulewidth}{0.04cm}
	\label{tab:simParam}
	\begin{tabular}{c c c | c c c}
		\hline
		Parameter & units & Value & Parameter & units & Value\\
		\hline
		$\TLWR_{\UB}$ & Celsius & 15   & $\TCWRUB$    & $^\circ\mathrm{C}$ & 40 \\
		$S_{\LB}$    & N/A     & 0.05 & $S_{\UB}$    & N/A     & 0.95 \\
		$\MTES$     & kg &  $1.413 \times 10^5$  &   $\TLWR_{\UB}$ & $\degree$C & 15 \\
		$\TApp_{\LB}$   & Celsius &  0 &  $\TApp_{\UB}$ & $^\circ\mathrm{C}$ &  40 \\
		$\TRan_{\LB}$   & Celsius &  2.2 &  $\TRan_{\UB}$ & $^\circ\mathrm{C}$ & 22.2 \\
          $\mDotCOND_{\noml}$   & kg/s &  47.44 &  $\TChWS_{\UB}$ & $^\circ\mathrm{C}$ & 10 \\
          $\TCONDWS_{\LB}$   & Celsius &  15 &  $\TCONDWS_{\UB}$ & $^\circ\mathrm{C}$ & 40 \\
          $\TCONDWR_{\noml}$   & Celsius &  35 &  $\TCONDWR_{\UB}$ & $^\circ\mathrm{C}$ & 40\\
		\hline
	\end{tabular}
\end{table}

\section{Closed Loop Simulation Example}\label{sec:sim}
%\rd{$\TChWS_{\set}$ was fixed at a constant $7\degree$C}
 The proposed model has been used for training RL controllers for a
 \plant\ in~\cite{GuoPhdThesis:2023}. Due to lack of space and ease of exposition, here we present a simpler
 application of the model: closed-loop simulation results with a rule-based
controller. The controller uses real-time electricity price to update
the control command $u_k$ at every $k$ with a goal to reduce total
electricity cost and meet the load $\qDotL$. We omit the details of the controller due to lack of
space, the interested reader can find them in~\cite{GuoPhdThesis:2023}.

\begin{figure}[ht]
  \centering
            \subfigure[2 of 3][\label{fig:exog-1}]{	\includegraphics[width=0.99\columnwidth]{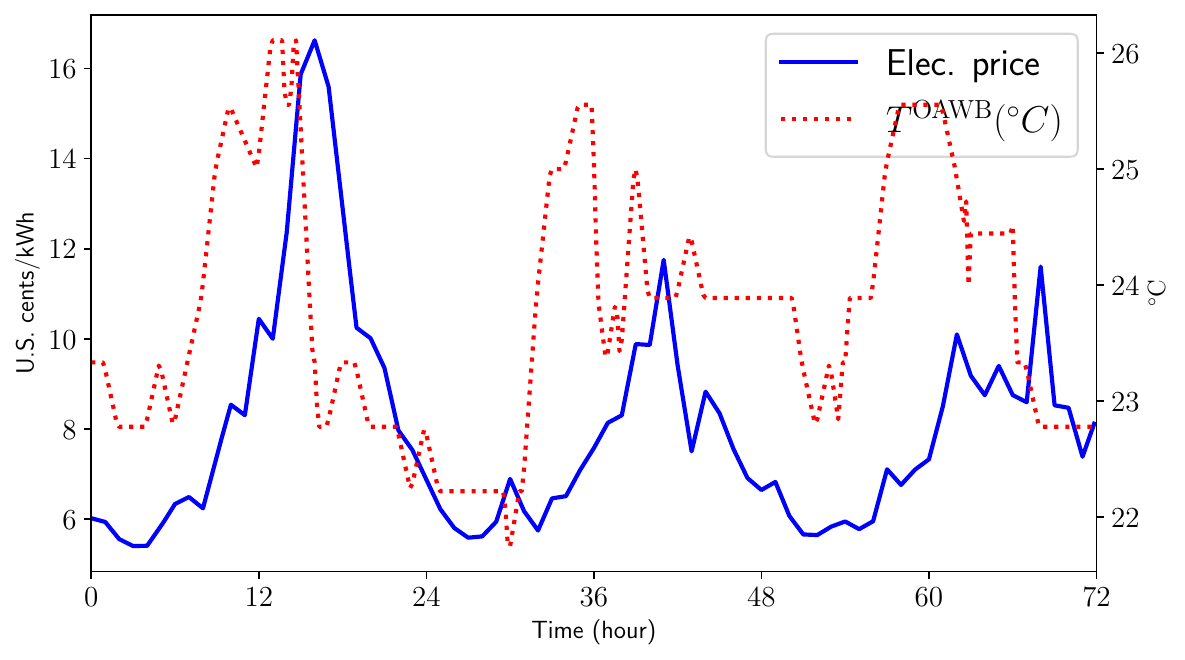}}
            \subfigure[2 of 3][\label{fig:exog-2}]{	\includegraphics[width=0.99\columnwidth]{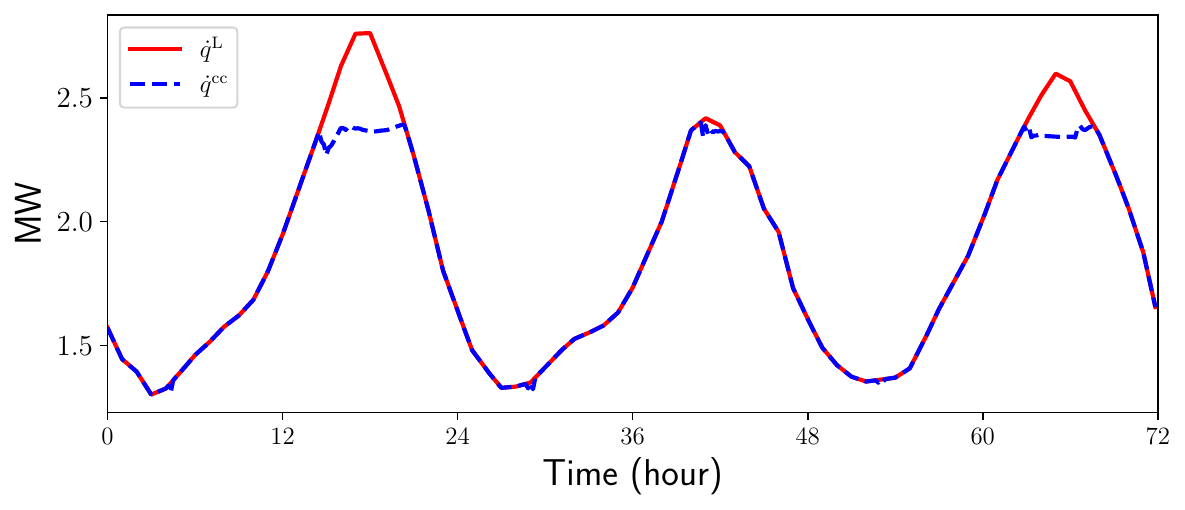}}
	\caption{Closed loop simulation: All three exogenous inputs $w^p$, and cooling provided $\dot{q}^{\cc}$.}
	\label{fig:exog}
\end{figure}

Fig.~\ref{fig:exog} shows the three exogenous signals used in the
simulations: OA wet bulb temperature, electricity price (top), and cooling demand $\qDotL$, cooling delivered $\qDotCC$ (bottom), over a three-day
period. Note that the user needs to provide these three exogenous inputs in order to perform simulations with the
\plant\ model and the rule based controller. The OA wet bulb
temperature is obtained from is obtained from the National Solar Radiation
Database~\url{https://nsrdb.nrel.gov}, for the location for
Gainesville, FL, USA. Electricity price is taken from
PJM~\cite{PJM:ElecPriceUrl}. The cooling load is chosen as a scaled version of PJM's
total electricity demand for the same period, since cooling demand and
electricity demand are correlated. The scaling is done to make the
maximum cooling load equal to 130\% of the total cooling capacity of the two chillers combined. These
closed loop simulations can be recreated by running the \matlab\ script
\texttt{simulateCCWP\_withRBC.m} in the codebase. 

\begin{figure}[ht]
	\includegraphics[width=0.98\columnwidth]{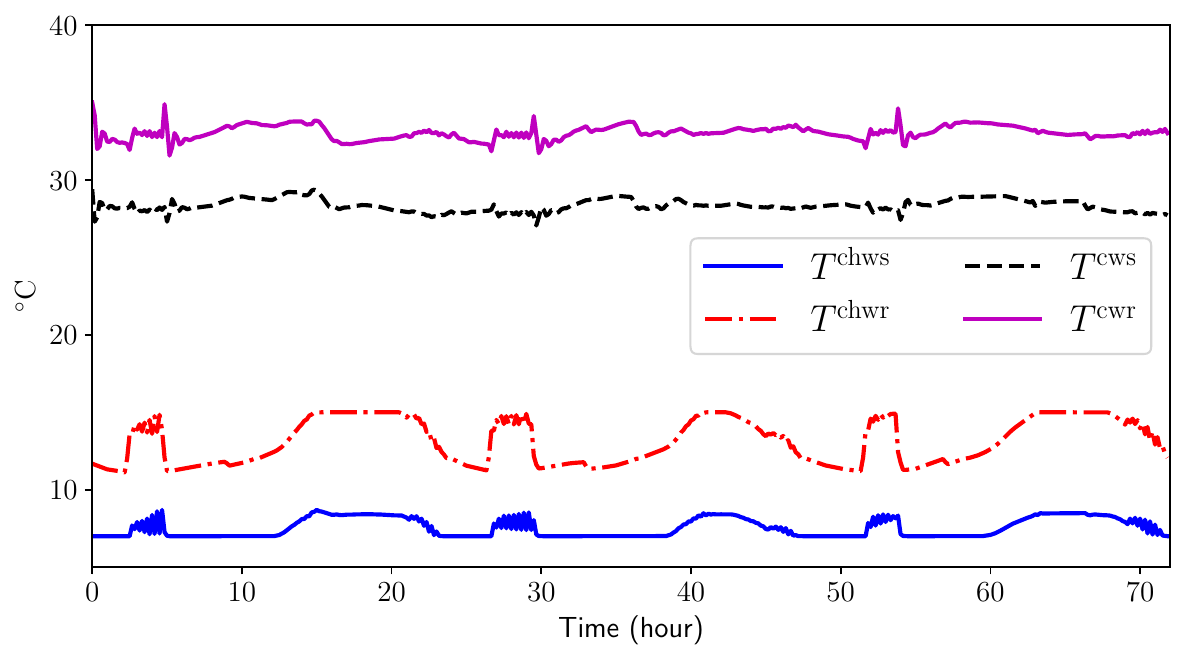}
	\caption{Closed loop simulation: Chilled water and cooling water temperatures.}
	\label{fig:CL-ChWCWTemp}
\end{figure}

We can see from Fig.~\ref{fig:exog-2} that the plant is able to meet the cooling load
for most, but not all, of the time. 
\begin{figure}[ht]
	\centering
	\includegraphics[width=0.98\columnwidth]{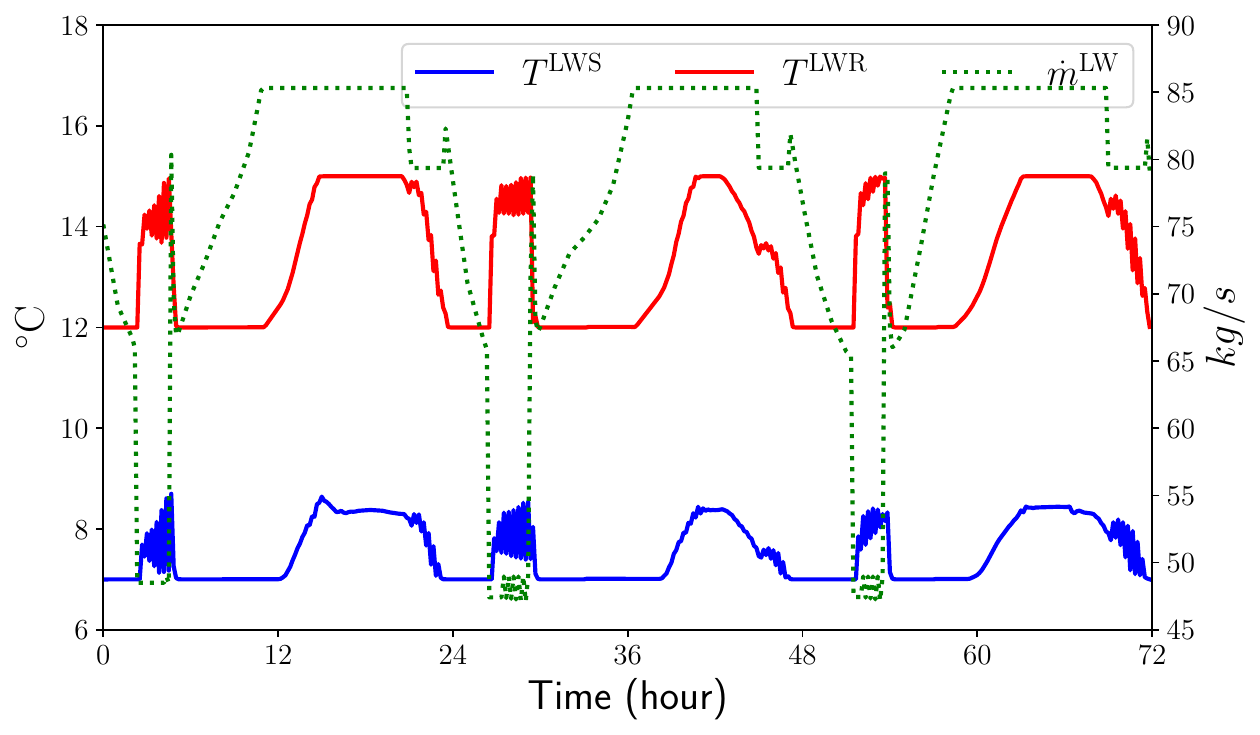}
	\caption{Closed loop simulation: Saturation of cooling coil heat exchange during periods of high load.}
	\label{fig:CL-CC}
\end{figure}
During periods when the cooling load exceeds cooling capacity, both the water flow
through the cooling coil and the temperature at its exit saturate to
their upper bounds; see Fig.~\ref{fig:CL-CC}. Fig.~\ref{fig:CL-ChWCWTemp} shows that the
various water temperatures are within bounds at all
times; the kind of unrealistic high temperatures seen in
Section~\ref{sec:OriginalChillerFail} never occurs. The oscillations observed at certain times is due to frequent
changes in the TES flow rate commands computed by the rule based
controller. The controller needs to be improved to make the closed
loop response smoother. 

The design COP of the two chillers used in the simulations are
  6.28 and 7.35. The plantwide COP, computed as $(\sum_{k=0}^{k_{\max}}\qDotCC_k)/(\sum_{k=0}^{k_{\max}}\Ptot_k)$, turns out to be
  5.84, i.e., 93-80\% of the chillers'. A plantwide COP that is
  between 60 - 80\% of chiller's COP is considered
  ``OK'' and above 80\% is considered
  ``good''~\cite{personalCommDavidBrooks:2023}. These numbers
  therefore give us confidence on both the controller and the \plant\ model.

  \section{Conclusions}\label{sec:conclusion}
   The proposed model has been used for training an RL
   controller for a \plant\ in~\cite{GuoPhdThesis:2023}. There are many improvements that can be made to the model and its
software implementation. As to the software, translating the codebase
from \matlab\ to Python or Julia will obviate the need for commercial
software with a license fee. Improvements are possible to the
mathematical model itself, such as  incorporating additional dynamic
elements and transport delays. %Our goal is to develop a model of a
                               %\plant\ that is useful for training
                               %learning-based controllers, not create a general purpose simulator.

Verification of the proposed \plant\ model against experimental data is not addressed in this paper. The accuracy of the proposed model is dependent on the accuracy of the constituent equipment models. It was shown in~\cite{ZhangFromBS:2022} that many existing and widely used chiller models for predicting COP can work well in laboratory conditions but provide poor predictions in field conditions. The component models we use are ultimately based on existing models in the literature and may suffer from similar issues. However, since the proposed \plant\ model is modular, improved component models can be incorporated in the future. 

% \appendix{Extra stuff}
% For Sept.2022, with BL controller, the cost is \$14467 for next-time model, and is \$14233 for current-time model; with RL(MLQ) controller, the cost is \$13581 for next-time model, and is \$13025 for current-time model.
\bibliographystyle{IEEEtran}

\small
% Generated by IEEEtran.bst, version: 1.14 (2015/08/26)

\end{document}